\documentclass[letter]{scrartcl}	

\usepackage{enumitem} %To customise lists.
\usepackage{setspace} % To format text line spacing.

\usepackage{bm} % To get bold math.
\usepackage{float} % To get floats
\usepackage[dvipsnames]{xcolor} % To get many colours.
\usepackage{amsmath, mathtools, amsthm, mathrsfs, amssymb} % MATH Packages
\usepackage[upright]{fourier} % To customise font.
\usepackage{secdot} % To put dots after section numbering.
\usepackage[export]{adjustbox} % To resize anything.
\usepackage[section]{placeins} % To place figures itextfrakn their section.
\usepackage[labelfont=bf,skip=10pt,labelsep=space,justification=centering]{caption} % To customise captions.
\usepackage{yfonts} % To get Gothic Math letters.
\usepackage[framemethod=tikz]{mdframed} % To get page-breakable proofs.
\usepackage{anyfontsize} % To avoid "substitution with differences" warning.
\usepackage{silence} % To avoid annoying unnecessary warnings.
\usepackage{cancel} % To cross some fancy things.
\usepackage[final]{microtype} % For the sake of perfection.
\usepackage[low-sup]{subdepth} % For subscripts and upscripts to be homogeneous.
\usepackage{nicefrac} % To get nice fractions in text mode.
\usepackage{multicol} % To get two columns.
\usepackage{titling} % To customise titling.
\usepackage{geometry} % To custom layout.
\usepackage{mathalfa} % To custom fonts in math environment.
\usepackage{lscape} % To get a landscape page.
\usepackage{multirow} % To get several columns / lines in tables.
\usepackage{csquotes} % To make sure that my Bibliography loads correctly.
\usepackage{tabularx} % For fancy tables.
\usepackage{etoolbox} % For re-sizing and stuff.
\usepackage{pifont} % For cross and check marks.
\usepackage{fontawesome} % For fancy symbols.
\usepackage{subcaption} % To get subcaption
\usepackage{setspace}
\usepackage{dsfont}
\usepackage{bbm}
\usetikzlibrary{decorations.pathreplacing,calligraphy} %Curly brackets in tikz

\usepackage[bottom]{footmisc}

\newcommand{\rvline}{\hspace*{-\arraycolsep}\vline\hspace*{-\arraycolsep}}

\usepackage[square,numbers]{natbib}
\patchcmd\subequations{\theparentequation\alph{equation}}{\theparentequation\thechildequation}{}{\error}
\def\thechildequation{\Alph{equation}}

\makeatletter

\newtheoremstyle{mystyle0}{0}{}{\setlength\parindent{0pt}}{}{\scshape}{}{ }{\thmname{#1} \thmnumber{#2}.\thmnote{#3}}
\theoremstyle{mystyle0}
\newmdtheoremenv[skipabove=20pt,skipbelow=10pt,innerleftmargin=0.0cm,innerrightmargin=0.0cm,innertopmargin=0.0cm,innerbottommargin=0.0cm,backgroundcolor=White!10,linecolor=Black!60,linewidth=0.5mm,hidealllines=true,nobreak=false,roundcorner=2.5pt,startinnercode={\deffootnotemark{}\deffootnote{0pt}{0pt}{}}]{proposition}{Proposition}[]

\newmdtheoremenv[skipabove=20pt,skipbelow=10pt,innerleftmargin=0.0cm,innerrightmargin=0.0cm,innertopmargin=0.0cm,innerbottommargin=0.0cm,backgroundcolor=White!10,linecolor=Black!60,linewidth=0.5mm,hidealllines=true,nobreak=false,roundcorner=2.5pt,startinnercode={\deffootnotemark{}\deffootnote{0pt}{0pt}{}}]{lemma}{Lemma}[]

\newmdtheoremenv[skipabove=20pt,skipbelow=10pt,innerleftmargin=0.0cm,innerrightmargin=0.0cm,innertopmargin=0.0cm,innerbottommargin=0.0cm,backgroundcolor=White!10,linecolor=Black!60,linewidth=0.5mm,hidealllines=true,nobreak=false,roundcorner=2.5pt,startinnercode={\deffootnotemark{}\deffootnote{0pt}{0pt}{}}]{corollary}{Corollary}[]

\newtheoremstyle{mystyle2}{0}{}{\setlength\parindent{0pt}}{}{\scshape}{}{ }{\thmname{#1} \thmnumber{#2} --- \thmnote{#3}.}
\theoremstyle{mystyle2}
\newmdtheoremenv[skipabove=20pt,skipbelow=20pt,innerleftmargin=0.0cm,innerrightmargin=0.0cm,innertopmargin=0.0cm,innerbottommargin=0.0cm,linecolor=Black!60,linewidth=0.5mm,hidealllines=true,nobreak=false,roundcorner=2.5pt,startinnercode={\deffootnotemark{}\deffootnote{0pt}{0pt}{}}]{definition}{Definition}[]

\newtheoremstyle{mystyle3}{0}{}{\setlength\parindent{40pt}}{}{\scshape}{}{ }{\thmname{#1} \thmnumber{#2}.\thmnote{#3}}
\theoremstyle{mystyle3}
\newmdtheoremenv[skipabove=20pt,skipbelow=20pt,innerleftmargin=0.0cm,innerrightmargin=0.0cm,innertopmargin=0.0cm,innerbottommargin=0.0cm,linewidth=0.5mm,backgroundcolor=White,nobreak=false,hidealllines=true,roundcorner=5pt,linecolor=Black!60,startinnercode={\deffootnotemark{}\deffootnote{0pt}{0pt}{}}]{example}{Example}

\newtheoremstyle{mystyle1}{0}{}{\setlength\parindent{40pt}}{}{\scshape}{}{ }{\thmname{#1} \thmnumber{#2}.\thmnote{#3}}
\theoremstyle{mystyle1}

\newmdtheoremenv[skipabove=10pt,skipbelow=20pt,innerleftmargin=0.0cm,innerrightmargin=0.0cm,innertopmargin=0.0cm,innerbottommargin=0.0cm,hidealllines=true,backgroundcolor=White!10,nobreak=false,linewidth=0.5mm,linecolor=Gray!70,roundcorner=2.5pt,startinnercode={\deffootnotemark{}\deffootnote{0pt}{0pt}{}}]{proof}{Proof}[]

\AtEndEnvironment{proof}{\null\hfill$\blacksquare$}

\newtheoremstyle{mystylex}{0}{}{\setlength\parindent{40pt}}{}{\scshape}{}{ }{\thmname{#1} \thmnumber{#2} --- \thmnote{#3}.}
\theoremstyle{mystylex}
\newmdtheoremenv[skipabove=10pt,skipbelow=20pt,innerleftmargin=0.0cm,innerrightmargin=0.0cm,innertopmargin=0.0cm,innerbottommargin=0.0cm,hidealllines=true,backgroundcolor=White!10,nobreak=false,linewidth=0.5mm,linecolor=Gray!70,roundcorner=2.5pt,startinnercode={\deffootnotemark{}\deffootnote{0pt}{0pt}{}}]{uproof}{Proof}[]
\AtEndEnvironment{uproof}{\null\hfill$\blacksquare$}

\makeatother

\usepackage[colorlinks=true,urlcolor=blue,linkcolor=black]{hyperref}
\usepackage[noabbrev,]{cleveref}
\hypersetup{allcolors=Blue}



\newtheorem{assumption}{Assumption}

\DeclareSymbolFont{mathup}{T1}{fut\mathfamilyextension}{m}{n}

\DeclareMathSymbol{a}{\mathalpha}{mathup}{`a}
\DeclareMathSymbol{b}{\mathalpha}{mathup}{`b}
\DeclareMathSymbol{c}{\mathalpha}{mathup}{`c}
\DeclareMathSymbol{d}{\mathalpha}{mathup}{`d}
\DeclareMathSymbol{e}{\mathalpha}{mathup}{`e}
\DeclareMathSymbol{f}{\mathalpha}{mathup}{`f}
\DeclareMathSymbol{g}{\mathalpha}{mathup}{`g}
\DeclareMathSymbol{h}{\mathalpha}{mathup}{`h}
\DeclareMathSymbol{i}{\mathalpha}{mathup}{`i}
\DeclareMathSymbol{j}{\mathalpha}{mathup}{`j}
\DeclareMathSymbol{k}{\mathalpha}{mathup}{`k}
\DeclareMathSymbol{l}{\mathalpha}{mathup}{`l}
\DeclareMathSymbol{m}{\mathalpha}{mathup}{`m}
\DeclareMathSymbol{n}{\mathalpha}{mathup}{`n}
\DeclareMathSymbol{o}{\mathalpha}{mathup}{`o}
\DeclareMathSymbol{p}{\mathalpha}{mathup}{`p}
\DeclareMathSymbol{q}{\mathalpha}{mathup}{`q}
\DeclareMathSymbol{r}{\mathalpha}{mathup}{`r}
\DeclareMathSymbol{s}{\mathalpha}{mathup}{`s}
\DeclareMathSymbol{t}{\mathalpha}{mathup}{`t}
\DeclareMathSymbol{u}{\mathalpha}{mathup}{`u}
\DeclareMathSymbol{v}{\mathalpha}{mathup}{`v}
\DeclareMathSymbol{w}{\mathalpha}{mathup}{`w}
\DeclareMathSymbol{x}{\mathalpha}{mathup}{`x}
\DeclareMathSymbol{y}{\mathalpha}{mathup}{`y}
\DeclareMathSymbol{z}{\mathalpha}{mathup}{`z}

\let\oGamma\Gamma
\renewcommand{\Gamma}{\mathchoice{\raisebox{0pt}{\scalebox{1}{$\oGamma$}}}{\oGamma}{\oGamma}{\oGamma}}
\let\oDelta\Delta
\renewcommand{\Delta}{\mathchoice{\raisebox{0pt}{\scalebox{1}{$\oDelta$}}}{\oDelta}{\oDelta}{\oDelta}}

\let\oTheta\Theta
\renewcommand{\Theta}{\mathchoice{\raisebox{0pt}{\scalebox{1}{$\oTheta$}}}{\oTheta}{\oTheta}{\oTheta}}

\let\oLambda\Lambda
\renewcommand{\Lambda}{\mathchoice{\raisebox{0pt}{\scalebox{1}{$\oLambda$}}}{\oLambda}{\oLambda}{\oLambda}}

\let\oPi\Pi
\renewcommand{\Pi}{\mathchoice{\raisebox{0pt}{\scalebox{1}{$\oPi$}}}{\oPi}{\oPi}{\oPi}}

\let\oSigma\Sigma
\renewcommand{\Sigma}{\mathchoice{\raisebox{0pt}{\scalebox{1}{$\oSigma$}}}{\oSigma}{\oSigma}{\oSigma}}
\let\oUpsilon\Upsilon
\renewcommand{\Upsilon}{\mathchoice{\raisebox{0pt}{\scalebox{1}{$\oUpsilon$}}}{\oUpsilon}{\oUpsilon}{\oUpsilon}}
\let\oPhi\Phi
\renewcommand{\Phi}{\mathchoice{\raisebox{0pt}{\scalebox{1}{$\oPhi$}}}{\oPhi}{\oPhi}{\oPhi}}
\let\oPsi\Psi
\renewcommand{\Psi}{\mathchoice{\raisebox{0pt}{\scalebox{1}{$\oPsi$}}}{\oPsi}{\oPsi}{\oPsi}}
\let\oOmega\Omega
\renewcommand{\Omega}{\mathchoice{\raisebox{0pt}{\scalebox{1}{$\oOmega$}}}{\oOmega}{\oOmega}{\oOmega}}

\let\existstemp\exists
\let\nexiststemp\nexists
\let\foralltemp\forall
\renewcommand*{\exists}{\mkern4mu\existstemp\mkern0mu}
\renewcommand*{\nexists}{\mkern0mu\nexiststemp\mkern4mu}
\renewcommand*{\forall}{\mkern4mu\foralltemp\mkern0mu}

\renewcommand{\prod}{\mathchoice{\oprod}{\raisebox{1.17pt}{\scalebox{0.775}{$\oprod$}}}{\oprod}{\oprod}}
\let\obigcap\bigcap
\renewcommand{\bigcap}{\mathchoice{\obigcap}{\raisebox{1.17pt}{\scalebox{0.775}{$\obigcap$}}}{\obigcap}{\obigcap}}
\let\obigcup\bigcup
\renewcommand{\bigcup}{\mathchoice{\obigcup}{\raisebox{1.17pt}{\scalebox{0.775}{$\obigcup$}}}{\obigcup}{\obigcup}}

\let\originalleft\left
\let\originalright\right
\renewcommand{\left}{\mathopen{}\mathclose\bgroup\originalleft}
\renewcommand{\right}{\aftergroup\egroup\originalright}

\makeatletter
\newcommand{\vast}{\bBigg@{3}}
\newcommand{\Vast}{\bBigg@{4}}
\newcommand{\VVast}{\bBigg@{5}}
\newcommand{\VVVast}{\bBigg@{6}}
\makeatother

\setkomafont{disposition}{\normalfont}
\setkomafont{sectioning}{\scshape}
\addtokomafont{section}{\centering}
\addtokomafont{subsection}{\centering}
\sectiondot{subsection}

\setlist{after=\vspace{3.5pt}}
\setenumerate{leftmargin=*,labelsep=32.5pt,itemsep=0pt,topsep=2.5pt}
\setitemize{leftmargin=17.5pt,labelsep=5pt,itemsep=0pt,topsep=2.5pt}



\makeatletter
\newcommand{\onetagright}{\tagsleft@false}
\makeatother

\usepackage{titling}
\begin{document}

\title{MORAL HAZARD WITH NETWORK EFFECTS\\  \Large(JOB MARKET PAPER)}
\author{\Large MARC CLAVERIA MAYOL\thanks{Stony Brook University; claveriamayol.marc@stonybrook.edu. \newline I thank my advisor Mihai Manea for his guidance and support. I also thank Eran Shmaya, Ting Liu, Pradeep Dubey, Ben Golub, In\'{e}s Macho-Stadler, Ran Shorrer, Takahiro Moriya, H\'{e}ctor Hermida-Rivera, and Robert Millard for their useful comments.}  \\[1cm]}
\date{\today \\ \href{https://www.marcclaveriamayol.com/research}{Latest version.}}

\spacing{1.3}
\maketitle

\vspace{.5cm}

\spacing{1.3}

 \begin{abstract} 
 	I study a moral hazard problem between a principal and multiple agents who experience positive peer effects represented by a (weighted) network. Under the optimal linear contract, the principal provides high-powered incentives to central agents in the network in order to exploit the larger incentive spillovers such agents create. The analysis reveals a novel measure of network centrality that captures rich channels of direct and indirect incentive spillovers and characterizes the optimal contract and its induced equilibrium efforts. The notion of centrality relevant for incentive spillovers in the model emphasizes the role of pairs of agents who link to common neighbors in the network. This characterization leads to a measure of marginal network effects and identifies the agents whom the principal targets with stronger incentives in response to the addition (or strengthening) of a link. When the principal can position agents with heterogeneous costs of effort in the network, the principal prefers to place low-cost agents in central positions. The results shed light on how firms can increase productivity through corporate culture, office layout, and social interactions.
\end{abstract}

	

\section{Introduction}

Network architecture has important implications for the optimal provision of incentives. In the presence of peer effects, the actions of influential workers affect those of their co-workers. For example, \citet{MasMoretti} find that the introduction of a highly productive cashier in a supermarket boosts the productivity of others on the same shift by $1\%$. They observe that these effects are experienced by workers in their line of vision and that they decrease with the distance between cashiers. Variations of such effects are found in top-management teams (\citet{HayesOyer}), physicians within health maintenance organizations (\citet{Gaynor}) or in flower packing workers from different ethnicities in Kenya (\citet{Kenya}). \citet{Bandiera} provide a review of empirical evidence for such effects and conclude that their impact on productivity is economically relevant. In this paper, I study how social benefits from interactions with peers shape a manager's incentive provision to a group of agents producing individual outputs.

% \footnote{\citet{BandieraNegative} look at workers experiencing a negative version of such effects when picking fruit due to the design of the incentive scheme at the farm.}

I study a moral hazard problem in a multi-agent setting when these are embedded into a (directed) network of interactions. I build upon a well-established moral hazard model (\citet{HolmMil,HolmMil91}) in which agents have constant absolute risk aversion (CARA) preferences, and the principal offers a linear contract that specifies for each agent fixed and variable (or performance-related) terms of the compensation that depends on the agent's observed output.\footnote{For a comprehensive introduction to the model and its various applications, please refer to \citet{Bolton} or \citet{Macho2018}.} The network captures benefits from the quality of the work environment, reduction in the cost of effort derived from interactions with peers, or other social preferences and status concerns. These network effects enter the CARA utility function as a product of the agent's effort and the effort of each, the intensity of the directed link, and a parameter measuring the global strength of peer effects. The network is assumed to be exogenous and known to the principal. Given the contracts offered by the principal, agents simultaneously choose efforts at a quadratic cost of effort. Each agent's output depends on individual effort and a random shock; shocks have independent normal distributions.

% These network effects enter the CARA utility function as a product of the efforts of the pair of agents, the intensity of the link between the two of them, and a parameter measuring the universal strength of network effects  OR An agent's network effects enter the CARA utility function as the sum of effort products for each pair and the intensity of that directed link, and a parameter measuring the global strength of network effects. 

%Agents simultaneously choose their efforts which depend on that of their neighbors. An agent's output depends on effort and an exogenous random variable. Thus, the principal designs contracts for each agent that only depend on individual outputs which, unlike the efforts, are observed. However, this does not mean all agents receive the same compensation since influential agents are more valuable to the principal as sources of spillovers. 

\begin{figure}[t]
	\centering
	\includegraphics[width=175pt]{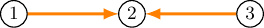}
	\caption{A directed line network with three agents.}
	\label{3line}
\end{figure}

Better connected agents enhance the benefits of exerting effort for their neighbors. The principal optimally exploits incentive spillovers by providing high-powered incentives to central agents. To develop intuition, consider the network in Figure \ref{3line} in which agents 1 and 3 are affected by agent 2. The links capture how agents 1 and 3 care about agent 2's effort. That is, an increase in 2's effort would lead to an increase in agents 1 and 3's marginal return of effort which in turn would incentivize them to increase their own efforts. The principal recognizes this effect and understands that incentives provided to agent 2 will not only increase 2's effort but also those of 1 and 3. Thus, the principal can take advantage of incentive spillovers, provide high-powered incentives to agent 2, and potentially contract with agents 1 and 3 for lower compensation to increase profits. 

I first characterize the optimal contract and its induced levels of effort for an arbitrary network. For every specification of the linear contracts, the principal induces a linear-quadratic network game, as in \citet{Ballester}. In this network game, equilibrium efforts are uniquely characterized and closely tied to agents' Bonacich centralities. The principal is able to affect this network game by choosing compensations for each agent. Anticipating the behavior of agents in the network game, the principal maximizes expected profits, i.e., the sum of individual outputs minus total compensations, taking into account agents' participation constraints. The principal utilizes the fixed term of the compensation to guarantee each agent obtains their reservation utility and the performance-related term to provide direct and indirect incentives. The notion of centrality relevant for incentive spillovers in the model emphasizes the role of pairs of agents who link to common neighbors in the network. In fact, performance-related compensations depend on a vector of agents' total direct and indirect influences as well as on a matrix of bilateral common influences on others. In addition to the obvious direct channel of influence, the analysis reveals new channels that the principal can exploit to augment incentive spillovers in all possible directions.  

% For an agent to be a source of incentive spillovers for the principal it must be that some other agent cares about that agent's effort. Therefore, we can express the influence that an agent has on others by the total discounted walks in the network that end at that agent. However, this measure alone is not enough for the principal to decide how to allocate incentives to agents, since 

Next, I derive a measure of marginal network effects that captures the impact of considering global peer effects on the principal's profits. This measure boils down to a weighted sum of links, which suggests that the principal prefers dense networks to iterate on network effects in all directions as much as possible. 

I also conduct comparative statics exercises to examine how the optimal performance-related compensations and the equilibrium levels of effort depend on changes in the network and the strength of network effects. I find that adding (or strengthening) a directed link between two agents, e.g., from agent $i$ to agent $j$, results in non-negative effects on optimal performance-related compensations and efforts. In particular, it (strictly) increases the performance-related compensation of agent $j$ as well as that of any agent that (i) is weakly connected to agent $j$,\footnote{Agent $i$ is weakly connected to agent $j$ if they are connected when considering the undirected network.} and (ii) has some influence on at least one agent. Furthermore, it also (strictly) increases the effort of agents $i$, $j$, and that of any other agent weakly connected to them. I also find that increasing the strength of network effects increases the optimal performance-related compensation and equilibrium effort of all agents with at least one link. 

Lastly, I consider an application in which the network is interpreted as an office layout and workers have heterogeneous costs of effort. For example, an open-office concept allows workers to constantly interact with peers, and so a complete network would capture such a setting. At the opposite extreme, an empty network would represent a situation in which workers work in isolation. One more example is that of a line of cashiers in a supermarket as in \citet{MasMoretti}. In all these cases, the principal is exogenously constrained by the office space. When the principal faces high-cost agents in central positions, high-powered incentives can be targeted to non-central but low-cost agents when network effects are small. However, as network effects increase, the principal switches to targeting the high-cost but central agents to leverage incentive spillover. Moreover, if the principal can position agents in the network, low-cost agents are positioned in the most central positions in the network.

%An important observation is that, from the perspective of an agent, the complementarities between agents' efforts and the flow of incentives operate in opposite directions. A link from $i$ to some $j$ allows the former to enjoy complementarities with the latter. On the other hand, links from $j$ to $i$, make $i$ an attractive agent in the eyes of the principal since incentives allocated to $i$ will spill over to $j$. \textcolor{red}{What's the point of saying this?}

\medskip 

\textbf{Literature.} This paper belongs to the literature on network games. This literature has been growing at a steady pace since the seminal work by \citet{Ballester} and \citet{GGJVY}.\footnote{For surveys on the literature on network games see \citet{BK} or \citet{JZ}.} In recent years, there has been some research that supplements games with complementarities with an entity that can affect payoffs in the game. For instance, \citet{GGG} consider a planner maximizing welfare by choosing the vector of benefits from an agent's action, which is independent of others' actions. The counterpart of the planner in this paper is the principal. In both papers, the planner makes a decision that affects the Nash equilibrium of the game the agents play. While in \citet{GGG} the decision amounts to spreading a budget among the agents, in the current setting the principal has to decide the fixed and performance-related terms of the compensation while taking into account the agents' participation constraints.

In contemporaneous and complementary work, \citet{Golub1} consider a moral hazard relationship between a principal and a group of agents producing a joint output that the principal observes. The same complementarity term appears in both papers; for them, it enters the production function, while in this paper it enters the agents' utility function. Another difference is that their complementarities are symmetric while in this paper complementarities can be asymmetric and are critical for understanding which agents receive higher compensation. In both papers, it turns out that the principal prefers dense networks, although the mechanisms and the nature of indirect spillovers in the two models are different. In a closely related working paper, \citet{Milan} consider the same model and characterize optimal linear contracts when there is joint production. They also consider situations in which the principal can only offer type-based contracts, where types represent job descriptions. \citet{Shi} studies a model with team production in which agents who share a link can help each other to reduce the cost of effort. \citet{Belhaj} study a contracting problem when agents are embedded into a network of (symmetric) complementarities when the principal wants to maximize (or minimize) aggregate effort and can only target one agent to contract with.

Lastly, this paper is also related to the literature on moral hazard in teams initiated by \citet{Holm}. As mentioned above, I build upon the model of \citet{HolmMil}. They study incentive provision to a single agent over time and show that a linear incentive scheme is optimal.\footnote{It is generally not true that linear contracts are optimal in static settings (see \citet{Mirrlees}). \citet{Carroll} shows that linear contracts are optimal in the case with limited liability, risk neutrality, and when the principal is uncertain about the actions available to the agent.} This model has been extended to situations in which agents' outputs are correlated and the principal uses relative performance evaluation, which makes an agent's wage contingent on the output of some other agent(s).\footnote{See \citet{Bolton} or \citet{Macho2018}.} For example, \citet{Bartling} conducts a conceptually related exercise analyzing a multi-agent contracting problem when outputs are correlated and agents have preferences related to inequality or status concerns. In such cases, relative performance evaluation is valuable as it reduces agents' risk exposure. However, in this paper, if the principal makes an agent's compensation contingent on the output of some other agent(s), the principal would only be raising the agent's risk exposure.

\medskip 

The rest of the paper is organized as follows. Section 2 presents a H\"{o}lmstrom-Milgrom moral hazard model with multiple agents and introduces a network of peer effects. Section 3.1 presents the optimal contract in the absence of a network. Sections 3.2 and 3.3 characterize the equilibrium levels of effort for each agent and the optimal linear contracts for the principal for an arbitrary network. In Section 4, I derive a measure of marginal network effects. Section 5 presents comparative statics. In Section 6, I consider agents with heterogeneous costs of effort and investigate how the principal would like to allocate agents to nodes in the network. Section 7 concludes. Proofs omitted in the main body of the paper appear in the Appendix.

\section{Model}

Consider a moral hazard relationship between a principal and a set of agents $\mathcal{N}=\{ 1, 2, \dots, n \}$. Each agent $i\in\mathcal{N}$ simultaneously chooses a level of effort, $a_{i}\geq 0$, that results in an individual noisy output according to
\begin{equation}
	q_{i} = a_{i} + \varepsilon_{i}.
\end{equation}
The random variables $(\varepsilon_{i})_{i \in \mathcal{N}}$ are assumed to be independently and normally distributed with mean zero and variance $\sigma^{2}$. Denote an action profile by $\mathbf{a}=(a_{1},\dots.a_{n})$. Bold lower-case letters are used for vectors $\mathbf{b}\in\mathbb{R}^{n}$, and bold upper-case letters, $\mathbf{B}$, for matrices. 

As in \citet{HolmMil}, agents are risk averse with CARA preferences. The novelty is that agents' payoffs also depend on a weighted network that captures peer effects described by a matrix $\mathbf{G}$. The $(i,j)$th entry of the adjacency matrix is denoted by $g_{ij}\geq 0$ and represents the intensity of the interaction of agent $i$ with agent $j$.\footnote{In general,  $g_{ij}\neq g_{ji}$. That is, the matrix $\textbf{G}$ need not be symmetric.} Each agent $i\in\mathcal{N}$ has a utility function given by 
\begin{equation}
		u_{i}(\mathbf{a},\mathbf{G})= -exp\left(-\eta \left[w_{i} - \frac{1}{2}ca_{i}^{2} + \beta \sum\limits_{j\in\mathcal{N}} g_{ij}a_{i}a_{j}\right]\right), \label{utilityfunct}
\end{equation}
where $\eta$ denotes the coefficient of absolute risk aversion, $w_{i}$ the agent's compensation, $c\geq 0$ the cost of effort, and $\beta \geq 0$ is the global strength of network effects. The last term captures the complementarities between an agent and that of his neighbors. In particular, $\sum\nolimits_{j\in\mathcal{N}} g_{ij}a_{j}$ captures the positive impact of neighbors' actions on $i$'s marginal benefit of effort. By convention, $g_{ii}=0$ for every $i\in\mathcal{N}$.

% Footnote on the case in which $\beta=0$

The principal is risk neutral and offers linear contracts. Agent $i$'s compensation is
\begin{equation}
w_{i} = z_{i} + v_{i}q_{i},	
\end{equation}
where $z_{i}$ is a fixed term of the compensation, and $v_{i}$ is a variable or performance-related compensation coefficient. A linear contract is a pair $(\mathbf{z},\mathbf{v})$ that specifies for each $i\in\mathcal{N}$ the fixed and performance-related terms of the compensation. 

The principal publicly announces a contract $(\mathbf{z},\mathbf{v})$, that is, each agent's compensation is common knowledge. The focus is on situations in which, given the contractual terms, agents play the Nash equilibrium of the induced game, and every agent $i$ finds the resulting outcome acceptable given their reservation wage $\underline{w}_{i}$. Therefore, the principal solves the following optimization problem:
%\bigskip 

%\textbf{Timing.} First, the principal designs and proposes a contract to each agent. Then, each $i\in\mathcal{N}$ decides whether to accept or reject the contract. After that, agents simultaneously choose their effort levels taking into account network effects. Lastly, individual outputs are realized and payments are executed. 

%\bigskip 

\begin{align}
		\max_{(a_{i},z_{i},v_{i})_{i=1}^{N}} & \quad  E\left(\sum_{i=1}^{n}q_{i} - \sum_{i=1}^{N}w_{i} \right) \tag{\text{P1}} \\
		\text{s.t.} & \quad E\left( -e^{-\eta [w_{i} - \frac{1}{2}c a^2_i + \beta \sum_{j\in N} g_{ij}a_{i}a_{j}]} \right) \geq -e^{-\eta\underline{w}_{i}} & \forall i =1,\dots,n  \tag{\text{PC}}\label{eq:PC} \\
		&  \quad a_{i} \in \text{arg}\max_{\hat{a}_{i}} E\left( -e^{-\eta [w_{i} - \frac{1}{2}c \hat{a}_{i}^{2} + \beta \sum_{j\in N} g_{ij}\hat{a}_{i}a_{j}]} \right) & \forall i =1,\dots,n   \tag{\text{ICC}}\label{eq:ICC}
\end{align}
where \eqref{eq:PC} is agent $i$'s participation constraint, which says that the principal has to guarantee each agent their reservation utility for the agent to accept the contract, and \eqref{eq:ICC} is the incentive compatibility constraint which imposes that the efforts chosen by the agents are part of a Nash equilibrium of the network game. The principal needs to anticipate agents' decisions for any given contract and, then, choose the optimal contract.

\section{Optimal Contracts and Efforts}

In this section, I first analyze the case with no network effects. Next, I solve for the Nash equilibrium of the network game in which agents simultaneously choose efforts for any specification of linear contracts.  The section concludes with the characterization of the optimal contract for the principal. 

\subsection{The Case With No Network Effects}

The optimal contract and equilibrium effort when there is an empty network ($g_{ij}=0$ for all $i,j\in\mathcal{N}$) applies as the baseline for the case with no network effects, i.e., $\beta=0$, as well as for cases with non-empty networks in which some agents work in isolation. In these cases, agents' contracts are completely independent from one another and so the problem reduces to the case of a principal contracting with a single agent. In the textbook treatment of this model, I simply present the optimal contract and its induced effort. 

For all $i\in \mathcal{N}$, the optimal contract for the principal and the agent's effort are given by
\begin{equation}
\begin{aligned}
		\left(z_{i}^{\emptyset}  = \underline{w}_{i} - a^{\emptyset}_{i}v^{\emptyset}_{i} + \frac{1}{2}c(a^{\emptyset}_{i})^{2} + \eta \frac{\sigma^{2}}{2}(v^{\emptyset}_{i})^{2},\quad  v_{i}^{\emptyset}  = \frac{1}{1+c\eta\sigma^{2}}\right)\quad \text{and} \quad 	a^{\emptyset}_{i}   = \frac{v_{i}^{\emptyset}}{c}.
	\label{optimalcontractempty}
\end{aligned}
\end{equation}

Note that the optimal performance-related compensation and the equilibrium effort are decreasing in the cost of effort $c$, the coefficient of absolute risk aversion $\eta$, and the variance of the output shock $\sigma^{2}$. 

\subsection{Nash Equilibrium in the Network Game}

Let $\mathbf{G}$ be an arbitrary network and suppose that all agents participate in the relationship. For any given linear contract $(\mathbf{z},\mathbf{v})$, each agent $i\in\mathcal{N}$ chooses a level of effort to maximize expected utility. Standard results leveraging that the output shocks $\varepsilon_{i}$ are normally distributed with mean zero and variance $\sigma^{2}$ can be used to show that maximizing the CARA utility function \cref{utilityfunct} is equivalent to maximizing the agent's certainty equivalent (CE), which consists of the expected compensation, benefits from network effects, cost of effort, and risk exposure. That is, each agent maximizes the (linear-quadratic) CE given by
\begin{equation*}
		CE = z_{i} + a_{i}\left( v_{i} + \beta \sum_{j\in \mathcal{N}}g_{ij}a_{j} \right) - \frac{1}{2}c a_{i}^{2} - \eta \frac{\sigma^{2}}{2}v_{i}^{2}.  
\end{equation*}

Taking the first-order condition with respect to agent $i$'s action and rearranging, agent $i$'s best response can be written as
\begin{equation}
	a_{i}^{*} = \frac{v_{i}+\beta\sum_{j\in \mathcal{N}} g_{ij} a_{j}}{c}.
	\label{Agenti_BR}
\end{equation}
Agent $i$'s best response reveals that the principal can affect an agent's effort in two ways: (i) by directly increasing agent $i$'s performance-related compensation and/or (ii) by increasing the performance-related compensation of $i$'s neighbors, which incentivizes $i$ to increase effort due to complementarities with other's efforts. Writing the system of first-order conditions for all $i\in\mathcal{N}$ in matrix form, the necessary condition for the effort profile $\textbf{a}^{*}$ to be a Nash equilibrium is
\begin{equation}
	c\left[ \mathbf{I} -\lambda\mathbf{G}\right] \mathbf{a^{*}} = \mathbf{v},
	\label{NE_matrixform}
\end{equation}
where $\mathbf{I}$ denotes the identity matrix and $\lambda=\beta/c$.

\medskip

\begin{assumption}
	The spectral radius of $\lambda \mathbf{G}$ is less than 1.\footnote{Recall that the spectral radius of a matrix $\mathbf{A}$ is the eigenvalue with the largest absolute value.}
\end{assumption}

Assumption 1 can be interpreted as bounding the strength of network effects. Under this assumption, the matrix $[\mathbf{I}-\lambda\mathbf{G}]$ is invertible and its inverse has non-negative entries.\footnote{See \citet{Debreu}.} In this case, \eqref{NE_matrixform} is necessary and sufficient for determining the Nash equilibrium action profile $\mathbf{a}^{*}$ of the network game, and there exists a unique Nash equilibrium given by 
\begin{equation}
	\mathbf{a}^{*} = \frac{1}{c}\left[ \mathbf{I} - \lambda\mathbf{G} \right]^{-1} \mathbf{v}.
	\label{NE_unique}
\end{equation}

Let $\mathbf{M}(\mathbf{G},\lambda) = [\mathbf{I} - \lambda\mathbf{G}]^{-1} = \sum_{k=0}^{\infty}\lambda^{k}\mathbf{G}^{k}$ and denote its $(i,j)$th entry by $m_{ij}(\mathbf{G},\lambda)$, or simply by $m_{ij}$ when reference to the original adjacency matrix $\mathbf{G}$ or the parameter $\lambda$ is not crucial. As \citet{Ballester} explain, $m_{ij}$ accounts for the total weight of all walks in $\mathbf{G}$ that start in $i$ and end at $j$, where walks of length $k$ are weighted by $\lambda^{k}$.\footnote{A \emph{walk} from $i$ to $j$ of length $k$ is a sequence $\langle i_{0},\dots, i_{k}\rangle$ of agents such that $i_{0}=i$, $i_{k}=j$ and $g_{i_{p}i_{p+1}}>0$ for all $0\leq p\leq k-1$ such that $i_{p}\neq i_{p+1}$. Note that a walk is different from a path because it does not require all agents in the sequence to be distinct. When the network is unweighted, i.e., $g_{ij}=\{0,1\}$ for all $i,j\in\mathcal{N}$, the $(i,j)$th entry of $G^{k}$ is the number of walks from $i$ to $j$ of length $k$.} Note that by definition $m_{ii}\geq 1$. When all agents receive identical performance-related compensation, they show that the equilibrium profile of actions is proportional to the vector of Bonacich centralities, defined by $\mathbf{b}(\mathbf{G},\lambda) = [ \mathbf{I} - \lambda\mathbf{G} ]^{-1}\cdot \mathbf{1}$.

%% --------------------------------------------------------------------------------------------- %%
%% --------------------------------------------------------------------------------------------- %%
%% --------------------------------------------------------------------------------------------- %%
%% --------------------------------------------------------------------------------------------- %%
%% --------------------------------------------------------------------------------------------- %%
%% --------------------------------------- SUBSECTION ------------------------------------------ %%
%% --------------------------------------------------------------------------------------------- %%
%% --------------------------------------------------------------------------------------------- %%
%% --------------------------------------------------------------------------------------------- %%
%% --------------------------------------------------------------------------------------------- %%

\subsection{Optimal Contracts} 

Anticipating agents' behavior in the network game, the principal chooses the contract $(\mathbf{z},\mathbf{v})$ that maximizes expected payoff. Note that the participation constraints of all agents will bind by standard moral hazard arguments. To see this, suppose the participation constraint of agent $i$ is not binding. The principal can reduce agent $i$'s fixed compensation while ensuring that the agent still accepts the contract. This reduction in compensation increases the principal's payoff. Consequently, the principal can continue to decrease $z_{i}$ until the participation constraint holds with equality. This illustrates the role of the fixed compensation in the model --- it serves as a mechanism to minimize the principal's costs while simultaneously ensuring that agents receive their guaranteed reservation utility. Therefore, by plugging an agent's payoff in the network game into the participation constraint, the fixed compensation of agent $i\in\mathcal{N}$ can be written as 
\begin{equation*}
	z_{i} = \underline{w}_{i} + \eta \frac{\sigma^{2}}{2}v_{i} - \frac{c}{2}(a_{i}^{*})^2.
\end{equation*}

Next, substituting this expression into the principal's objective function, the maximization is reduced to
\begin{align*}
		\max_{(v_{i})_{i=1}^{n}} & \quad  \sum_{i\in N} a_{i}^{*}(1-v_{i}) - \sum_{i\in N} \underline{w}_{i} + \frac{1}{2}c \sum_{i\in N} (a_{i}^{*})^{2} - \eta \frac{\sigma^{2}}{2}\sum_{i\in N} v_{i}^{2}.
\end{align*}
whereby \eqref{NE_unique}, the Nash equilibrium effort of agent $i\in\mathcal{N}$ is given by 
\begin{equation*}
	a_{i}^{*} = \frac{1}{c}\sum_{j=1}^{n}m_{ij}v_{j}. 
\end{equation*}
      
Taking first-order conditions with respect to $v_{i}$ and rearranging terms we obtain
\begin{equation*}
	v_{i} = \frac{1}{c\eta\sigma^{2}}\left[\sum_{j=1}^{n}m_{ji} - \sum_{j=1}^{n}m_{ij}v_{j}-\sum_{j=1}^{n}m_{ji}v_{j} + \sum_{r=1}^{n}m_{ri}\sum_{s=1}^{n}m_{rs}v_{s}\right].
\end{equation*} 

Note that for $r=i$ the fourth term in the brackets above can be grouped with the second term to obtain $(m_{ii}-1)\sum_{j=1}^{n}m_{ij}v_{j}$, which is greater or equal to zero since $m_{ii}\geq  1$ by definition. Similarly, for $j\neq i$, the third term in the brackets above can be grouped with the fourth term for $r=s=j$ as $\sum_{j=1}^{n}(m_{jj}-1)m_{ji}v_{j}$, which again is non-negative since $m_{jj}\geq 1$. Therefore, we can rewrite these equations for all $i\in\mathcal{N}$ as follows: 

\begin{equation}
	v_{i} = \frac{1}{m_{ii}+c\eta\sigma^{2}}\left[\sum_{j=1}^{n}m_{ji} + (m_{ii}-1)\sum_{j=1}^{n} m_{ij}v_{j}  + \sum_{j\neq i}(m_{jj}-1)m_{ji}v_{j} + \sum_{r\neq \{i,j\}}^{n}m_{ri}\sum_{s=1}^{n}m_{rs}v_{s}  \right].
	\label{OC_BRi(vj)}
\end{equation}

It is useful to interpret \eqref{OC_BRi(vj)} as the principal's ``best-response of $v_{i}$ to $v_{j}$, $j\neq i$.'' The first term in the brackets of \eqref{OC_BRi(vj)} represents the total influence of agent $i$ on all agents, including himself. Note that $m_{ji}$ can be interpreted as the \emph{direct influence that agent $i$ has on agent $j$}, since $j$ experiences network effects with $i$ only if there is a walk from $j$ to $i$. Thus, the more influence $i$ has on others, the higher his compensation. The second term compares the benefits from iterated network effects via $i$'s self-cycles, i.e., $m_{ii}\sum_{j=1}^{n}m_{ij}v_{j}$, with ``saving money'' motives via incentives spillovers from $i$'s neighbors that make $i$ cheaper to incentivize, i.e., $-\sum_{j=1}^{n}m_{ij}v_{j}$. Since $m_{ii}\geq 1$, iterated network effects dominate ``saving money'' motives, strictly if there is at least one self-cycle for $i$. The third term captures other agents' iterated effects via agent $i$'s influence on them. Again, for each $j\neq i$ that is influenced by agent $i$, these effects iterate through cycles starting and ending at $j$. The fourth term captures common influences of agents $i$ and $s$ on other agents.

%The second term captures agent $i$'s iterated effects of incentive spillovers from others, including himself. Note that for network effects to iterate there must be some agent $k$ for which $m_{ik}$ and $m_{ki}$ are positive. In other words, this term captures effects iterated via cycles in the network that start and end at agent $i$.  

The system of first-order conditions can be written in matrix form as \footnote{All steps to reach this expression are shown in the appendix.}
	\begin{equation}
		\left[ \mathbf{I} - \delta \left( \mathbf{M}(\mathbf{G},\lambda)\cdot\mathbf{G} \right)^{T}\cdot\mathbf{M}(\mathbf{G},\lambda)\cdot\mathbf{G} \right] \mathbf{v}=\frac{1}{1+c\eta\sigma^{2}}\boldsymbol{\alpha}, 
		\label{OC_FOC_MatrixForm2}
	\end{equation}
where $\delta = \lambda^{2}/(1+c\eta\sigma^{2})$ and $\boldsymbol{\alpha}$ is a column vector where the $i$th entry is $\sum_{j=1}^{n}m_{ji}$. This is a necessary condition for $\mathbf{v}$ to be the vector of optimal performance-related compensation. 

Remember that agent $i$'s Bonacich centrality is $b_{i}(\mathbf{G},\lambda)=\sum_{j=1}^{n}m_{ij}(\mathbf{G},\lambda)$. We have that $\alpha_{i}=\sum_{j=1}^{n} m_{ji}$, and so the vector $\boldsymbol{\alpha}$ is the transpose of the vector of Bonacich centralities, i.e., $\boldsymbol{\alpha}=(\mathbf{M}(\mathbf{G},\lambda))^{T}\cdot \mathbf{1}^{T} = \mathbf{b}^{T}(\mathbf{G},\lambda)$. The $i$th entry measures agent $i$'s total influence on other agents measured by the total weight of walks that end at $i$. The matrix $(\mathbf{M}(\mathbf{G},\lambda)\cdot\mathbf{G})$ is essentially capturing the same information as $\mathbf{M}(\mathbf{G},\lambda)$, ignoring walks of length zero. To see this,  write $\mathbf{M}(\mathbf{G},\lambda)\cdot\mathbf{G} = (\sum_{k=0}^{\infty}\lambda^{k}\mathbf{G}^{k})\cdot\mathbf{G}=\sum_{k=0}^{\infty}\lambda^{k}\mathbf{G}^{k+1}$.  Denote entry $(i,j)$th of $(\mathbf{M}(\mathbf{G},\lambda)\cdot\mathbf{G})$ by $\overline{m}_{ij}$. The product $(\mathbf{M}(\mathbf{G},\lambda)\cdot\mathbf{G})^{T}\cdot\mathbf{M}(\mathbf{G},\lambda)\cdot\mathbf{G}$ is a symmetric matrix\footnote{For a matrix $\mathbf{A}$, the products $\mathbf{A}\mathbf{A}^{T}$ and $\mathbf{A}^{T}\mathbf{A}$ give a square and symmetric matrix. In our case $\mathbf{A}=(\mathbf{M}\cdot\mathbf{G})$.} where the entry $(i,j)$ can be written as $\sum_{r=1}^{n}\overline{m}_{ri}\cdot \overline{m}_{rj}$, and represents the total common influence of $i$ and $j$ on all agents, including themselves. \medskip 

\begin{assumption}
	The spectral radius of the matrix $\delta (\mathbf{M}(\mathbf{G},\lambda)\cdot\mathbf{G} )^{T}\cdot\mathbf{M}(\mathbf{G},\lambda)\cdot\mathbf{G}$ is less than 1. 
\end{assumption} 

With this second assumption, the following matrix is well-defined:
\begin{equation*}
	\mathbf{W}(\mathbf{G}, \delta, \lambda)= [\mathbf{I} - \delta (\mathbf{M}(\mathbf{G},\lambda)\cdot\mathbf{G})^{T}\cdot\mathbf{M}(\mathbf{G},\lambda)\cdot\mathbf{G} ]^{-1} = \sum_{k=0}^{\infty} \delta^{k}((\mathbf{M}(\mathbf{G},\lambda)\cdot\mathbf{G})^{T}\cdot\mathbf{M}(\mathbf{G},\lambda)\cdot\mathbf{G})^{k}.
\end{equation*}

The entry $(i,j)$ of $\mathbf{W}(\mathbf{G}, \delta, \lambda)$ represents the total weight of the common influence of agents $i$ and $j$ on others as well as on each other. That is, $w_{ij}>0$ when both $i$ and $j$ influence a common agent, even if they do not influence each other. Generally, the $k$th power of the matrix $(\mathbf{M}\mathbf{G})^{T}\mathbf{M}\mathbf{G})^k$ keeps track of the common influence of two agents through a series of $k$ other agents with whom they have common influence. The parameter $\delta$ discounts joint influences through longer sequences of common influence. By definition, as was the case for $\mathbf{M}(\mathbf{G},\lambda)$, $w_{ii}\geq 1$ for all $i\in\mathcal{N}$. \medskip 

\begin{figure}[ht]
	\centering
	\includegraphics[width=300pt]{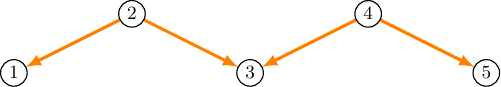}
	\caption{A directed network with five agents.}
	\label{5agents}
\end{figure}

\textbf{Example.} To develop intuition for the symmetric matrix $\mathbf{W}(\mathbf{G}, \delta, \lambda)$, consider the network in \cref{5agents} with five agents. The only (directed) links are from agent 2 to 1 and 3, and from agent 4 to 3 and 5. That is, agent 1 influences 2's effort, agent 5 influences 4's effort, and agent 3 influences both 2's and 4's efforts.

The only zero entries on $\mathbf{W}(\mathbf{G}, \delta, \lambda)$ are in the second and fourth rows and columns, except the diagonal entries which are equal to 1. This is because there are no links to agents 2 and 4, and so they do not influence others' actions. Thus, the optimal performance-related compensation for agents 2 and 4 is $v^{\emptyset}$, as described in section 3.1. Entries $w_{13}=w_{31}$ of $\mathbf{W}(\mathbf{G}, \delta, \lambda)$ are positive because agents 1 and 3 directly influence agent 2. To see this, note that for $k=1$, the $(1,3)$ (or $(3,1)$) entry of the product $((\mathbf{M}(\mathbf{G},\lambda)\cdot\mathbf{G})^{T}\cdot\mathbf{M}(\mathbf{G},\lambda)\cdot\mathbf{G})^{k}$ can be written as $\overline{m}_{21}\cdot \overline{m}_{23}$, which is positive since agent 2 is connected to both 1 and 3. Also, by \eqref{OC_BRi(vj)}, note that in this example, the second and third terms in the brackets will be zero for all $i\in \mathcal{N}$ since there are no cycles. Thus, the only terms that survive are the first term, represented by $\boldsymbol{\alpha}(\mathbf{G},\lambda)$, and the fourth term, which captures this direct common influence of 1 and 3 on 2. Analogously, entries $w_{35}=w_{53}$ are also positive. Entries $w_{15}=w_{51}$ are also positive due to agents 1 and 5 both influencing agents 2 and 4 via their direct common influence with agent 3. Again, this can be seen by investigating the product $((\mathbf{M}(\mathbf{G},\lambda)\cdot\mathbf{G})^{T}\cdot\mathbf{M}(\mathbf{G},\lambda)\cdot\mathbf{G})^{k}$ for $k=2$ which can be written as $(\overline{m}_{21}\cdot \overline{m}_{23})(\overline{m}_{43}\cdot \overline{m}_{45})$. \medskip 

% \footnote{Remember $\mathbf{W}(\mathbf{G}, \delta, \lambda)$ is a symmetric matrix, so $w_{31},w_{51}$ and $w_{53}$ are also positive.}

In addition to the obvious channels of direct influence captured by $\boldsymbol{\alpha}(\mathbf{G},\lambda)$, the symmetric matrix $\mathbf{W}(\mathbf{G}, \delta, \lambda)$ reveals new channels of common influences relevant to the principal when deciding incentives provision while balancing the agents' participation constraints.

% The principal is able to understand how iterated network effects depend on each agent and allocate incentives according to this less obvious channel. 

\begin{proposition}\label{OptimalContract} 
\textit{Suppose assumptions 1 and 2 hold. For any network $\mathbf{G}$, the (unique) optimal contract $(\mathbf{z}^{*},\mathbf{v}^{*})$ for the principal is given by:}
\begin{align}
	\mathbf{v}^{*} & = \frac{1}{1+c\eta\sigma^{2}}\mathbf{W}(\mathbf{G}, \delta, \lambda)\cdot \boldsymbol{\alpha}, \label{OptimalVarComp} \\
	\mathbf{z}^{*} & = \underline{w} + \eta \frac{\sigma^{2}}{2}\mathbf{v}^{*}\circ \mathbf{v}^{*}-\frac{c}{2}\mathbf{a}^{*}\circ \mathbf{a}^{*},\label{OptimalFixComp} 
\end{align}
\textit{where $\delta = \lambda^{2}/(1+c\eta\sigma^{2})$ and $\boldsymbol{\alpha}$ is the vector of direct influences.\footnotemark}

\textit{Given the optimal contracts $(\mathbf{z}^{*},\mathbf{v}^{*})$, the unique Nash equilibrium in the effort network game is uniquely characterized by} 
\begin{align}
	\mathbf{a}^{*} = \frac{1}{c}\left[\mathbf{I}-\frac{\beta}{c}\mathbf{G} \right]^{-1}\mathbf{v}^{*}.
	\label{OptimalEfforts} 
\end{align}
\end{proposition}

\footnotetext{Recall that the Hadamard product of two vectors is denoted by $\circ$, and is sometimes referred to as the entry-wise multiplication.}

The proposition shows that the performance-related component of agent $i$'s compensation is a linear combination of agents' total influences with weights given by $i$'s common influence with each $j\in \mathcal{N}$. That is, an agent receives higher incentives when holding a lot of direct and indirect influence on others and/or when sharing common influence with others who have a lot of direct and indirect influence. 

Note that for a completely isolated agent, the optimal contract and the induced equilibrium effort are described by those without a network as in Section 3.1.

\section{The Impact of the Network}

In this section, I derive a measure of network aggregate effects. 

\begin{proposition}\label{NetworkImpact}
	\textit{The derivative of the principal's profits with respect to the coefficient of network complementarities $\beta$, evaluated at $\beta=0$, is given by}
\begin{equation*}
	\frac{\partial \Pi}{\partial \beta}\Bigr|_{\substack{\beta=0}} = \kappa \sum_{i=1}^{n}\sum_{j=1}^{n}g_{ji}. 
\end{equation*}
\textit{where $\kappa = 1/(c^{2}(1+c\eta\sigma^{2})) + ((c-1)(1+c\eta\sigma^{2})/(c^{3}(c\eta\sigma^{2})^{2}) $ is a constant.}
\end{proposition}

Proposition \ref{NetworkImpact} provides a simple measure of network aggregate effects that is proportional to the total weight of all links in the network $\mathbf{G}$. This points to a preference for denser networks by the principal.

\section{Comparative Statics}

I now investigate how the optimal contract as well as the equilibrium level of efforts vary with changes in the network, the strength of network effects, and other parameters of the environment. The first result describes how the optimal contract and equilibrium efforts change when the network varies. Remember that two agents are weakly connected if they are connected when considering the undirected network. 

\begin{proposition}
	\textit{An increase in the weight $g_{ij}$ of the link from agent $i$ to agent $j$ (that does not violate assumptions 1 and 2) weakly increases the optimal performance-related compensations and equilibrium efforts for every agent. In particular, it strictly increases the
	\begin{itemize}
		\item[(i)]  optimal performance-related compensation of agent $j$ as well as of agents $k$ who are weakly connected to $j$ and have at least one in-link, i.e., there exists an $l\in\mathcal{N}$ such that $g_{lk}>0$;
		\item[(ii)] equilibrium level of efforts of agents $i$, $j$ as well as of agents $k$ who are weakly connected to $j$.
	\end{itemize}
	For all other agents, the increment in $g_{ij}$ does not affect the optimal contract or equilibrium effort.}
\end{proposition}

The proposition states that the optimal performance-related compensation and the equilibrium level of efforts will not decrease when the weight of some link is increased. Furthermore, it identifies agents for whom these will strictly increase. The proof is presented in the appendix.

\begin{figure}[t]
	\centering
	\includegraphics[width=350pt]{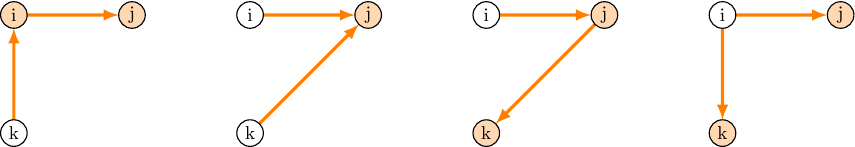}
	\caption[Short Title]{Some directed networks of three agents.  \par \small (Colored nodes represent agents for whom the optimal performance-related compensation strictly increases due to an increase in the weight $g_{ij}$ of the link from agent $i$ to agent $j$.)}
	\label{Fig.Prop3}
\end{figure}

Focus first on agents $i$ and $j$. An increase in the weight of the link from agent $i$ to $j$, makes the latter more influential on the former. By \eqref{Agenti_BR}, this leads to an increase in agent $i$'s effort since a larger $g_{ij}$ increases $i$'s marginal benefit from effort. Additionally, as agent $j$ gains direct influence on $i$, the principal responds by providing stronger incentives to $j$, capitalizing on the strengthened incentive spillovers. The increase on $v_{j}$ implies an increase in $a_{j}^{*}$ which, in turn, amplifies the increase in agent $i$'s effort. Thus, an increase in the weight $g_{ij}$ always leads to an increase in $a_{i}^{*}$, $v_{j}$, and $a_{j}^{*}$. Note that an increase in $v_{i}$ requires that there is at least one agent $k\in\mathcal{N}$ such that $g_{ki}>0$. Otherwise, incentives provided to agent $i$ cannot spill over to other agents and the principal sets $i$'s compensation equal to $v_{i}^{\emptyset}$, i.e., the compensation for an isolated agent described in section 3.1. This illustrates the second condition in part (i) of the proposition. For example, \cref{Fig.Prop3} shows different configurations for three-agent networks. When the weight $g_{ij}$ increases, agent $i$'s optimal performance-related compensation increases only for the first network thanks to the link from agent $k$ to $i$. 

Consider now an agent $k\in\mathcal{N}$ such that $g_{kj}>0$, as in the second network in \cref{Fig.Prop3}. In this case, agent $k$ is influenced by $j$'s effort. Agent $j$'s increased performance-related compensation and effort lead to an increase in $a_{k}^{*}$ through $k$'s best response. If instead $g_{jk}>0$, as in the third network in \cref{Fig.Prop3}, then $v_{k}$ would increase. In this case, since $k$ influences $j$ and $j$ influences $i$, then $k$ also has direct influence on $i$. An increase in the weight of the link from $i$ to $j$ leads to an increase in the indirect influence of $k$ on $i$, i.e., an increase in $m_{ik}$. As for $j$, the principal responds by providing stronger incentives to $k$ in order to capitalize on the enhanced incentive spillovers. Therefore, both $v_{k}$ and $a_{k}^{*}$ increase as a result of the increment in $g_{ij}$.

Next, consider an agent $k\in\mathcal{N}$ such that $g_{ik}>0$ and $g_{ki}=g_{kj}=g_{kj}=0$, as in the last network in \cref{Fig.Prop3}. Note that there is no walk from agent $j$ to agent $k$, or the other way around. However, because agent $i$ has directed links to both $k$ and $j$, the latter are weakly connected and influence agent $i$. As established above, the increase in the weight $g_{ij}$ leads to an increase in $a_{i}^{*}$. Therefore, relative to before the increase in $g_{ij}$, network effects from agent $k$ to agent $i$ are now amplified due to the larger $a_{i}^{*}$. This pushes the principal to increase $v_{k}$ to capitalize on the strengthened effects of incentive spillovers. That is, in this case, even though agent $k$'s influence on agent $i$ remains the same, the principal understands that the increase in the weight $g_{ij}$ of the link from $i$ to $j$ amplifies the effects from incentive spillovers from all agents influencing $i$. This effect is picked up by the last term in brackets of \eqref{OC_BRi(vj)} for agent $k$, i.e., common influences between agents $j$ and $k$ on $i$. The weakly connected condition in the proposition stems from the symmetry of $\mathbf{W}(\mathbf{G}, \delta, \lambda)$. This symmetry can also be seen from observing how \eqref{OC_BRi(vj)} considers incentive spillovers in all possible directions by considering walks in all directions.

To conclude with the intuition for Proposition 3, the equilibrium effort of all agents weakly connected to $j$ will increase due to the increased weight $g_{ij}$. Note that in all networks of \cref{Fig.Prop3}, the equilibrium efforts of all agents increase. In this model, higher performance-related compensation implies higher effort. Thus, all agents for whom the principal provides stronger incentives increase their efforts. Moreover, all agents who care about such agents, i.e., they have a link to an agent whose performance-related compensation increases, increase their own efforts due to the network effects affecting their marginal benefit from effort.

The next result explores the impact of varying the strength of network effects, $\beta$, and on the optimal variable compensation and the equilibrium level of efforts. 

\begin{proposition}\label{CS_ComplentarityParameters} 
\textit{An increase in the coefficient of network complementarities $\beta$ (that does not violate assumptions 1 and 2) results in increased optimal performance-related compensation and equilibrium efforts for all agents with at least one link in any direction.}
\end{proposition}

An increase in the strength of network effects leads to higher marginal benefits from effort when agents have at least one neighbor. Otherwise, an agent acts as if there is no network. Furthermore, the intensified network effects prompt the principal to leverage them further, increasing the performance-related compensation of all agents with some influence.

The final result in this section considers how optimal allocations depend on the rest of the parameters of the model, i.e., the cost of effort, the coefficient of absolute risk aversion, and the variance of the output shock. 

\begin{proposition}\label{CS_OtherParameters}
\textit{The optimal performance-related compensation $\mathbf{v}^{*}$ and the equilibrium level of efforts decrease with the agents' cost of effort $c$, the coefficient of risk aversion $\eta$ as well as with the variance of the output shock $\sigma^{2}$.}
\end{proposition}

This result mirrors comparative statics for the model without a network, linear contracts, and CARA risk preferences.

\section{Example with Heterogeneous Costs of Efforts}

Suppose a manager has a fixed working space and the network describes the layout of the offices with each node and social interactions in the work space capture the weights of links. Moreover, suppose the manager observes the social interactions between workers and each worker's cost of effort. \medskip 

\begin{figure}[hb]
	\centering
	\includegraphics[width=200pt]{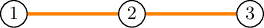}
	\caption{An undirected line network}
	\label{Ex6a}
\end{figure}

\textbf{Example.} Consider an undirected line network, as illustrated in \cref{Ex6a}, with $g_{12}=g_{21}=g_{23}=g_{32}=1$. Suppose $c_{2}=1$ and $c_{1}=c_{3}=0.5$. That is, agent 2 is a high-cost agent in the center of the line, and agents 1 and 3 are low-cost agents at the extremes. Normalize the following parameters:  $\eta =\sigma^{2} =1$ and $\underline{w}_{i}=0$ for all $i=\{ 1,2,3 \}$. How do the optimal performance-related compensations and equilibrium effort change as the coefficient on network effects increases? 

\begin{figure}[h!]
	\centering
	\includegraphics[width=400pt]{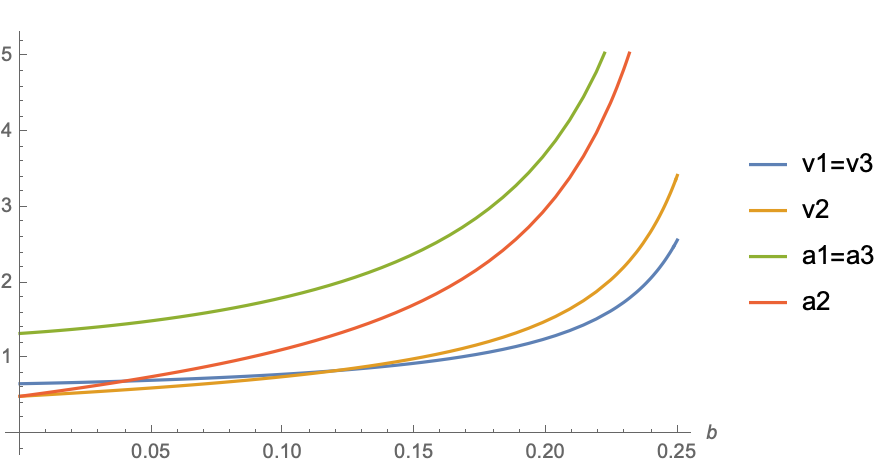}
	\caption{Optimal performance-related compensations and equilibrium efforts for a line network with a high-cost agent in the center.}
	\label{Ex1a}
\end{figure}

Optimal performance-related compensations and the equilibrium levels of effort are plotted in \cref{Ex1a} for different strengths of network effects $\beta$. For low values of $\beta$, the principal prefers to target low-cost agents with high-powered incentives. Despite these incentives spilling over to others in two ``steps'' --- from agent 1 to 2, and from 2 to 3 --- the lower cost of effort dominates iterated network effects. However, for $\beta$ high enough, the principal switches to targeting the central agent with high-powered incentives, even though agent 2 is a high-cost agent. That is, for strong network effects, the principal leverages the more efficient iteration of network effects that the central agent offers. Furthermore, note that, for all values of $\beta$, low-cost agents exert higher effort, despite receiving lower performance-related compensations when network effects are strong. 

\begin{figure}[t]
	\centering
	\includegraphics[width=400pt]{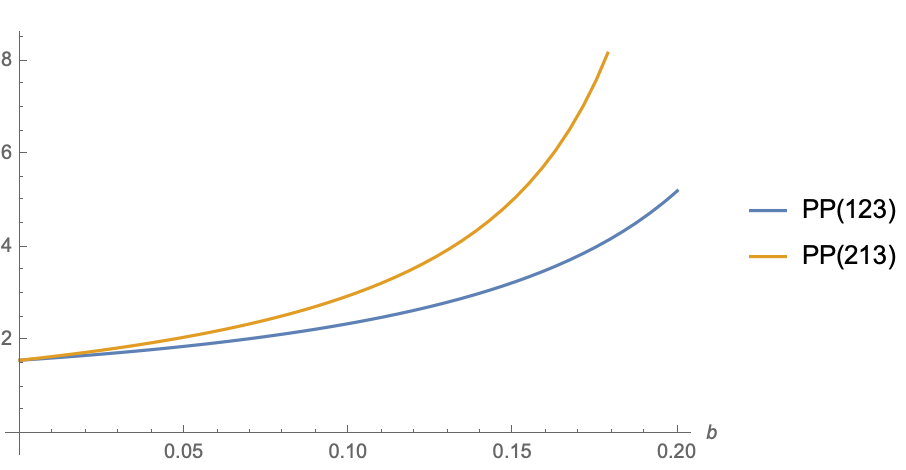}
	\caption{Principal's payoff for line networks with a high-cost or a low-cost agent central agent.}
	\label{Ex1b}
\end{figure}

Next, suppose the principal can position agents in different positions in the line network. In this case, the principal has two options: place either a low-cost or a high-cost agent in the center of the line. As we can see in \cref{Ex1b}, the principal prefers to allocate a low-cost agent in the center of the line for all possible values of $\beta$. 

% with a line network in which agent 2 is in the center. Then, for very small $\beta$ the principal gives higher incentives to the agents at the extreme. However, as $\beta$ increases, that is, as the network effect becomes bigger, the principal starts to provide higher incentives to the agent in the center, as illustrated in

\section{Conclusion}

I study a moral hazard relationship between a principal and a group of agents experiencing network effects that affect their marginal benefit from effort. In the model, agents have CARA risk preferences and produce individual output, and the principal offers linear contracts to agents. The principal can use more central agents to exploit incentive spillovers. This enables the principal to leverage network effects and achieve higher effort from agents at a lower cost.  

I first characterize the optimal contract for the principal and the induced Nash equilibrium in the agents' effort-provision network game. The analysis reveals that the optimal contract is tractable and, in addition to the obvious direct channels of influence among agents, depends on new channels of common influences between agents that the principal can exploit to iterate network effects throughout the network. Even though the kind of interactions I consider can be unidirectional, I find that what is relevant for the principal is the weakly connected network. This is captured by a novel symmetric matrix of bilateral common influences $\mathbf{W}(\mathbf{G},\delta,\lambda)$. I also derive a measure of network aggregate effects that is proportional to the sum of links in the network. Moreover, I analyze how the optimal contract and equilibrium efforts depend on the network and the strength of these network effects, and give conditions on network topology to identify agents for whom these strictly increase. 

Finally, motivated by a situation in which the principal can position agents in the network, i.e., offices, I analyze an example in which agents have heterogeneous costs of effort. The examples show that the provision of high-powered incentives depends on the strength of network effects and that, if possible, the principal prefers to position low-cost agents in highly central positions to iterate on network effects as much as possible. 

% -----------------------------------------------------------------------------------------------------------------------------------------------  %
% -----------------------------------------------------------------------------------------------------------------------------------------------  %
% ------------------------------------------------------------- REFERENCES ----------------------------------------------------------------------  %
% -----------------------------------------------------------------------------------------------------------------------------------------------  %
% -----------------------------------------------------------------------------------------------------------------------------------------------  %

%\renewcommand\arraystretch{1}
%\spacing{1}
%\emergencystretch=1em
%\printbibliography

\bibliography{bib}

\newpage

\section{Appendix}

\textit{Proof of the Claim in footnote 10.} I omit the dependency of the matrix $\mathbf{M}$ on the network $\mathbf{G}$ and the parameter $\beta/c$. The system of first-order conditions in \eqref{OC_BRi(vj)} can be written as:
\begin{equation*}
	\left[ \mathbf{I} + \frac{1}{c\eta\sigma^{2}}\left( \mathbf{M}+\mathbf{M}^{T} - \mathbf{M}^{T}\mathbf{M}\right) \right]\mathbf{v} = \frac{1}{c\eta\sigma^{2}}\boldsymbol{\alpha},
\end{equation*}
where $\boldsymbol{\alpha}$ is a column vector where the $i$th entry is $\sum_{j\in\mathcal{N}}m_{ji}$\footnote{Remember that $\mathbf{M}(\mathbf{G},\frac{\beta}{c})=(\mathbf{I}-\frac{\beta}{c}\mathbf{G})^{-1}$. To avoid cluttering, we omit the dependency on the adjacency matrix $\mathbf{G}$ and the complementarity and cost of effort parameters.}.

We rewrite the expression above, by using the following identity for inverse matrices:
	\begin{equation*}
		\left( \mathbf{I}-\mathbf{A}\right)^{-1} = \mathbf{I}+\left( \mathbf{I}-\mathbf{A}\right)^{-1}\mathbf{A},
	\end{equation*}
	
 and applying this identity on $\mathbf{M}$. That is, consider the product $\mathbf{M}^{T}\mathbf{M}$. Applying the identity on $\mathbf{M}$, we have
\begin{equation}
	\mathbf{M}^{T}\mathbf{M} = \mathbf{M}^T\left(\mathbf{I}+(\mathbf{I}-\frac{\beta}{c}\mathbf{G})^{-1}\frac{\beta}{c}\mathbf{G}\right) = \mathbf{M}^{T}+\frac{\beta}{c}\mathbf{M}^{T}\mathbf{M}\mathbf{G}.
	\label{A_L1_1}
\end{equation}
Similarly, taking the product $\mathbf{M}^{T}\mathbf{M}$ again but now applying the identity on $\mathbf{M}^{T}$, we have
\begin{equation}
	\mathbf{M}^{T}\mathbf{M} = \left(\mathbf{I}+(\mathbf{I}-\frac{\beta}{c}\mathbf{G})^{-1}\frac{\beta}{c}\mathbf{G}\right)^{T}\mathbf{M} = \left( \mathbf{I}+\frac{\beta}{c}\mathbf{G}^{T}(\mathbf{I}-\frac{
	\beta}{c}\mathbf{G})^{-T} \right)\mathbf{M} = \mathbf{M} + \frac{\beta}{c}\mathbf{G}^{T}\mathbf{M}^{T}\mathbf{M}, 
	\label{A_L1_2}
\end{equation}
where $\mathbf{M}^{-T}$ indicates the transpose of the inverse of $\mathbf{M}$.

We can now write 
\begin{equation*}
	\begin{aligned}
		\mathbf{M}+\mathbf{M}^{T}-\mathbf{M}^{T}\mathbf{M} & = \mathbf{M} - \frac{\beta}{c} \mathbf{M}^{T}\mathbf{M}\mathbf{G} & (\text{using \cref{A_L1_1}})\\
		& = \mathbf{M} - \frac{\beta}{c}(\mathbf{M}+\frac{\beta}{c}\mathbf{G}^{T}\mathbf{M}^{T}\mathbf{M})\mathbf{G} & (\text{using \cref{A_L1_2}})\\
		& = \mathbf{M} (\mathbf{I}-\frac{\beta}{c}\mathbf{G}) - \frac{\beta^{2}}{c^{2}}\mathbf{G}^{T}\mathbf{M}^{T}\mathbf{M}\mathbf{G}\\
		& = I - \frac{\beta^{2}}{c^{2}}\mathbf{G}^{T}\mathbf{M}^{T}\mathbf{M}\mathbf{G} & (\text{since} \hspace{0.15cm}\mathbf{M}=(\mathbf{I}-\frac{\beta}{c}\mathbf{G})^{-1})
	\end{aligned}
\end{equation*}

Therefore, we have that 
\begin{equation*}
	\left[ \mathbf{I} + \frac{1}{c\eta\sigma^{2}}\left( \mathbf{M}+\mathbf{M}^{T} - \mathbf{M}^{T}\mathbf{M}\right) \right] = \left[ (\frac{1+c\eta\sigma^{2}}{c\eta\sigma^{2}})\mathbf{I} - \frac{\beta^{2}}{c^{2}(c\eta\sigma^{2})} \mathbf{G}^{T}\mathbf{M}^{T}\mathbf{M}\mathbf{G}\right]
\end{equation*}

We can now rewrite the necessary condition for optimal performance-related compensation as
\begin{equation*}
	\left[ (\frac{1+c\eta\sigma^{2}}{c\eta\sigma^{2}})\mathbf{I} - \frac{\beta^{2}}{c^{2}(c\eta\sigma^{2})} \mathbf{G}^{T}\mathbf{M}^{T}\mathbf{M}\mathbf{G}\right]\mathbf{v}=\frac{1}{c\eta\sigma^{2}}\boldsymbol{\alpha}.
\end{equation*}

Finally, dividing both sides by $(1+c\eta\sigma^{2})/c\eta\sigma^{2}$ and noting that that the transpose of the product of two matrices $\mathbf{A}$ and $\mathbf{B}$ is equal to the transpose of $\mathbf{B}$ times the transpose of $\mathbf{A}$  we obtain
\begin{equation*}
		\left[ I - \delta \left( \mathbf{M}(\mathbf{G},\frac{\beta}{c})\cdot\mathbf{G} \right)^{T}\cdot\mathbf{M}(\mathbf{G},\frac{\beta}{c})\cdot\mathbf{G} \right] \mathbf{v}=\frac{1}{1+c\eta\sigma^{2}}\boldsymbol{\alpha}, 
	\end{equation*}
	where $\delta = \beta^{2}/(c^2(1+c\eta\sigma^{2})$. \hfill $\square$
	
%% ------------------------------------------------------------------------------------------------------ %%
%% ------------------------------------------------------------------------------------------------------ %%
%% ------------------------------------------------------------------------------------------------------ %%
%% --------------------------------- PROOF PROPOSITION 2 ------------------------------------------------ %%
%% ------------------------------------------------------------------------------------------------------ %%
%% ------------------------------------------------------------------------------------------------------ %%
%% ------------------------------------------------------------------------------------------------------ %%

\bigskip 
\noindent	
\textit{Proof of Proposition 2.} We can write the profits of the principal as follows
\begin{equation}
	\pi = (\mathbf{a}^{*})^{T}(\mathbf{1}^{T} - \mathbf{v}^{*})-\mathbf{1}^{T}\underline{\mathbf{w}} + \frac{c}{2}(\mathbf{a}^{*})^{T}\mathbf{a}^{*} - \eta\frac{\sigma^{2}}{2}(\mathbf{v}^{*})^{T}\mathbf{v}^{*}
	\label{PrincipalsProfits}
\end{equation}

We want to obtain the derivative of \eqref{PrincipalsProfits} with respect to the coefficient of network complementarities $\beta$ when evaluated at $\beta=0$. That is,
\begin{equation}
	\frac{\partial \Pi}{\partial \beta} = \mathbf{1}^{T}\frac{\partial \mathbf{a}^{*}}{\partial \beta} -  \frac{\partial (\mathbf{a}^{*})^{T}}{\partial \beta}\mathbf{v}^{*} - (\mathbf{a}^{*})^{T}\frac{\partial \mathbf{v}^{*}}{\partial \beta}+ \frac{c}{2}\left[\frac{\partial(\mathbf{a}^{*})^{T} }{\partial \beta}\mathbf{a}^{*} + (\mathbf{a}^{*})^{T}\frac{\partial \mathbf{a}^{*}}{\partial \beta}\right] - \eta\frac{\sigma^{2}}{2}\left[\frac{\partial (\mathbf{v}^{*})^{T}}{\partial \beta}\mathbf{v}^{*}+ (\mathbf{v}^{*})^{T}\frac{\partial\mathbf{v}^{*} }{\partial \beta}\right].
	\label{derivativeprincipalprofits}
\end{equation}

We first study the derivative of $\mathbf{v}^{*}$ with respect to $\beta$. It will be useful to use the equivalency in claim 1 and use the following expression for the optimal performance-related compensation vector
\begin{equation*}
	\mathbf{v}^{*} = \frac{1}{c\eta\sigma^{2}} \left[ \mathbf{I} + \frac{1}{c\eta\sigma^{2}}\left( \mathbf{M}(\beta)+\mathbf{M}(\beta)^{T} - \mathbf{M}(\beta )^{T}\mathbf{M}(\beta )\right) \right]^{-1}\boldsymbol{\alpha}(\beta) = \frac{1}{c\eta\sigma^{2}}\mathbf{\overline{M}}(\beta)\boldsymbol{\alpha}(\beta),
\end{equation*}
where we make explicit the dependency of the matrix $\mathbf{M}$ and the vector $\boldsymbol{\alpha}$ on the coefficient of network complementarities $\beta$\footnote{Both $\mathbf{M}$ and $\boldsymbol{\alpha}$ also depends on the adjacency matrix $\mathbf{G}$ and the cost of effort $c$.}. Taking the derivative with respect to $\beta$ we have
\begin{equation}
	\frac{\partial\mathbf{v}^{*} }{\partial \beta} = \frac{1}{c\eta\sigma^{2}}\left( \frac{\partial\overline{\mathbf{M}}(\beta) }{\partial \beta}\cdot \boldsymbol{\alpha}(\beta) + \overline{\mathbf{M}}(\beta) \cdot\frac{\partial \boldsymbol{\alpha}(\beta)}{\partial \beta}\right).
	\label{partialvbeta}
\end{equation}
The first term in the brackets involves the following derivative of the inverse of a matrix
\begin{equation}
	\begin{aligned}
	\frac{\partial\overline{\mathbf{M}}(\beta) }{\partial \beta} & = - \overline{\mathbf{M}}(\beta)\frac{\partial \left[ \mathbf{I} + \frac{1}{c\eta\sigma^{2}}\left( \mathbf{M}(\beta)+\mathbf{M}(\beta)^{T} - \mathbf{M}(\beta )^{T}\mathbf{M}(\beta )\right)\right]}{\partial \beta} \overline{\mathbf{M}}(\beta)\\
	& = \frac{1}{c\eta\sigma^{2}}\overline{\mathbf{M}}(\beta)\left[\frac{\partial \mathbf{M}(\beta)^{T}\mathbf{M}(\beta)}{\partial \beta}-\frac{\partial \mathbf{M}(\beta)}{\partial \beta}-\frac{\partial \mathbf{M}(\beta)^{T}}{\partial \beta} \right]\overline{\mathbf{M}}(\beta)\\
	& = \frac{1}{c\eta\sigma^{2}}\overline{\mathbf{M}}(\beta)\left[\left(\frac{\partial \mathbf{M}(\beta)^{T}}{\partial \beta}\mathbf{M}(\beta) + \mathbf{M}(\beta)^{T}\frac{\partial \mathbf{M}(\beta)}{\partial \beta}\right)-\frac{\partial \mathbf{M}(\beta)}{\partial \beta}-\frac{\partial \mathbf{M}(\beta)^{T}}{\partial \beta} \right]\overline{\mathbf{M}}(\beta).
	\end{aligned}
	\label{partialMbarbeta}
\end{equation}

Notice that 
\begin{equation}
	\begin{aligned}
	\frac{\partial \mathbf{M}(\beta)}{\partial \beta} & = -\mathbf{M}(\beta) \frac{\partial \left[ \mathbf{I}-\frac{\beta}{c}\mathbf{G} \right]}{\partial \beta} \mathbf{M}(\beta) = \mathbf{M}(\beta) \frac{\partial \frac{\beta}{c}\mathbf{G}}{\partial \beta}\mathbf{M}(\beta) = \frac{1}{c}\mathbf{M}(\beta)\mathbf{G}\mathbf{M}(\beta).
	\end{aligned}
	\label{A_DerivativeM_Beta}
\end{equation}

Using \eqref{A_DerivativeM_Beta} we can rewrite \eqref{partialMbarbeta} as 
\begin{equation*}
	\frac{\partial\overline{\mathbf{M}}}{\partial \beta} = \frac{1}{c(c\eta\sigma^{2})}\overline{\mathbf{M}}\left[\mathbf{M}(\beta)^{T}\mathbf{G}^{T}\mathbf{M}(\beta)^{T}\mathbf{M}(\beta) + \mathbf{M}(\beta)^{T}\mathbf{M}(\beta)\mathbf{G}\mathbf{M}(\beta)-\mathbf{M}(\beta)\mathbf{G}\mathbf{M}(\beta)-\mathbf{M}(\beta)^{T}\mathbf{G}^{T}\mathbf{M}(\beta)^{T} \right]\overline{\mathbf{M}}.
\end{equation*}

Notice that evaluating $\mathbf{M}(\beta)=[\mathbf{I}-\frac{\beta}{c}\mathbf{G}]^{-1}$ at $\beta=0$, we have that $\mathbf{M}(0)=\mathbf{I}$. Therefore, evaluating the last derivative at $\beta=0$, gives
\begin{equation*}
	\frac{\partial\overline{\mathbf{M}} }{\partial \beta} \Bigr|_{\substack{\beta=0}}= \frac{1}{c(c\eta\sigma^{2})}\overline{\mathbf{M}}(\beta)\left[\mathbf{G}^{T} + \mathbf{G}-\mathbf{G}-\mathbf{G}^{T} \right]\overline{\mathbf{M}}(\beta) = \mathbf{0}.
\end{equation*}

Focus now on the second term in brackets in \eqref{partialvbeta} and recall that $\boldsymbol{\alpha}$ is a column vector where the $i$th entry is $\sum_{j\in\mathcal{N}}m_{ji}(\beta)$. Taking the derivative of the $i$th entry of $\boldsymbol{\alpha}$, using $\eqref{A_DerivativeM_Beta}$, we have
\begin{equation}
	\frac{\partial \alpha_{i}}{\partial \beta} = \frac{\partial \sum_{j\in\mathcal{N}}m_{ji}(\beta)}{\partial \beta} = \sum_{j\in\mathcal{N}} \frac{\partial m_{ji}(\beta)}{\partial \beta} = \frac{1}{c}\sum_{j\in\mathcal{N}} \left( \mathbf{M}(\beta)\mathbf{G}\mathbf{M}(\beta) \right)_{ji} =\frac{1}{c} \sum_{j\in\mathcal{N}} \left( \sum_{l=1}^{n} \sum_{k=1}^{n}m_{jk}g_{kl}m_{li} \right).
	\label{A_Partialalpha_beta}
\end{equation}

Next, we evaluate this derivative at $\beta=0$ which is
\begin{equation*}
	\frac{\partial \alpha_{i}}{\partial \beta}\Biggr|_{\substack{\beta=0}} = \frac{1}{c}\sum_{j=1}^{n}g_{ji}
\end{equation*}

since all entries $m_{ij}$ such that $i\neq j$ will be zeros and when $i=j$ they will be ones since $\mathbf{M}(\beta=0)=\mathbf{I}$.

Evalutate \eqref{partialvbeta} at $\beta=0$
\begin{equation}
	\begin{aligned}
	\frac{\partial\mathbf{v} }{\partial \beta}\Bigr|_{\substack{\beta=0}} & = \frac{1}{c\eta\sigma^{2}}\left( \frac{\partial\overline{\mathbf{M}}(\beta) }{\partial \beta}\cdot \boldsymbol{\alpha}(\beta) + \overline{\mathbf{M}}(\beta) \cdot\frac{\partial \boldsymbol{\alpha}(\beta)}{\partial \beta}\right)\Bigr|_{\substack{\beta=0}} \\
	& = \frac{1}{c\eta\sigma^{2}} \left( \frac{\partial\overline{\mathbf{M}}(\beta) }{\partial \beta}\Bigr|_{\substack{\beta=0}}\cdot \boldsymbol{\alpha}(\beta)\Bigr|_{\substack{\beta=0}} + \overline{\mathbf{M}}(\beta)\Bigr|_{\substack{\beta=0}} \cdot\frac{\partial \boldsymbol{\alpha}(\beta)}{\partial \beta}\Bigr|_{\substack{\beta=0}}\right)\\
	& = \frac{1}{c\eta\sigma^{2}} \left( \left[ I + \frac{1}{c\eta\sigma^{2}} ( \mathbf{I}+\mathbf{I}^{T} - \mathbf{I}^{T}\cdot\mathbf{I} ) \right]^{-1} \cdot \frac{1}{c} \boldsymbol{\tilde{\alpha}} \right) \\
	& =\frac{1+c\eta\sigma^{2}}{c(c\eta\sigma^{2})^{2}}\boldsymbol{\tilde{\alpha}}
	\label{derivativevcomp_beta}
	\end{aligned}
\end{equation}
where $\tilde{\alpha}_{i} = \sum_{j=1}^{n}g_{ji}$. 

Next, we compute the derivative of the equilibrium vector of effort with respect to $\beta$. 
\begin{equation*}
	\begin{aligned}
		\frac{\partial \mathbf{a}^{*} }{\partial \beta}= \frac{1}{c}\left[ \frac{\partial \mathbf{M}}{\partial \beta}\cdot \mathbf{v}^{*} + \mathbf{M}\cdot \frac{\partial \mathbf{v}^{*}}{\partial \beta} \right] = \frac{1}{c}\left[ \frac{1}{c}\mathbf{M}\mathbf{G}\mathbf{M} \cdot \mathbf{v}^{*} + \mathbf{M}\cdot \frac{\partial \mathbf{v}^{*}}{\partial \beta} \right].
	\end{aligned}
\end{equation*}
Evaluating this derivative at $\beta=0$: 
\begin{equation}
	\begin{aligned}
		\frac{\partial \mathbf{a}^{*} }{\partial \beta}\Bigr|_{\substack{\beta=0}} & = \frac{1}{c} \left[ \frac{1}{c}\mathbf{I}\cdot\mathbf{G}\cdot\mathbf{I}\left(\mathbf{v}^{*}\Bigr|_{\substack{\beta=0}}\right) + \frac{1+c\eta\sigma^{2}}{c(c\eta\sigma^{2})^{2}}\boldsymbol{\tilde{\alpha}} \right] =  \frac{1}{c^{2}}\mathbf{G}\cdot \mathbf{v}^{\emptyset} + \frac{1+c\eta\sigma^{2}}{c(c\eta\sigma^{2})^{2}} \boldsymbol{\tilde{\alpha}},
		\label{derivativeefforts_beta}
	\end{aligned}
\end{equation}
where we have used that $(\mathbf{v}^{*}\bigr|_{\substack{\beta=0}})=\mathbf{v}^{\emptyset}$.

Now, we are equipped to go back to \cref{derivativeprincipalprofits} and evaluate that derivative at $\beta=0$ using \eqref{derivativevcomp_beta} and \eqref{derivativeefforts_beta}:
\begin{equation*}
	\begin{aligned}
	\frac{\partial \Pi}{\partial \beta}\Bigr|_{\substack{\beta=0}} & = \frac{1}{c^{2}}\mathbf{1}^{T}\mathbf{G}\mathbf{v}^{\emptyset} + \gamma \mathbf{1}^{T}\boldsymbol{\tilde{\alpha}} - \frac{1}{c^{2}}(\mathbf{v}^{\emptyset})^{T}\mathbf{G}\mathbf{v}^{\emptyset} - \gamma (\boldsymbol{\tilde{\alpha}})^{T}\mathbf{v}^{\emptyset} - \frac{\gamma}{c}(\mathbf{v}^{\emptyset})^{T}\boldsymbol{\tilde{\alpha}} \\ 
	& +\frac{c}{2} \left[ \frac{1}{c^{3}} (\mathbf{v}^{\emptyset})^{T}\mathbf{G}\mathbf{v}^{\emptyset} + \frac{\gamma}{c}(\boldsymbol{\tilde{\alpha}})^{T}\mathbf{v}^{\emptyset} + \frac{1}{c^{3}}(\mathbf{v}^{\emptyset})^{T}\mathbf{G}\mathbf{v}^{\emptyset}+\frac{\gamma}{c}(\mathbf{v}^{\emptyset})^{T}\boldsymbol{\tilde{\alpha}} \right] \\
	& - \eta \sigma^{2} \left[ \gamma (\boldsymbol{\tilde{\alpha}})^{T}\mathbf{v}^{\emptyset} + \gamma (\mathbf{v}^{\emptyset})^{T}\boldsymbol{\tilde{\alpha}} \right].
	\end{aligned}
\end{equation*}

Rearranging terms we obtain:
\begin{equation*}
	\begin{aligned}
	\frac{\partial \Pi}{\partial \beta}\Bigr|_{\substack{\beta=0}} &= \frac{1+c\eta\sigma^{2}}{c^{2}}(\mathbf{v}^{\emptyset})^{T}\mathbf{G}\mathbf{v}^{\emptyset} + \frac{(1+c\eta\sigma^{2})^{2}(c-1)}{c^{3}(c\eta\sigma^{2})^{2}}(\mathbf{v}^{\emptyset})^{T}\boldsymbol{\tilde{\alpha}} \\
	& = \frac{(1+c\eta\sigma^{2})[c(c\eta\sigma^{2})^{2}+(c-1)(1+c\eta\sigma^{2})^{2}]}{c^{3}(c\eta\sigma^{2})^{2}}(\mathbf{v}^{\emptyset})^{T}\mathbf{G}\mathbf{v}^{\emptyset},
	\end{aligned}
\end{equation*}
where in the last equality we use that $\boldsymbol{\tilde{\alpha}}=\mathbf{G}^{T}\cdot\mathbf{1}$ and group terms.

%% ------------------------------------------------------------------------------------------------------ %%
%% ------------------------------------------------------------------------------------------------------ %%
%% ------------------------------------------------------------------------------------------------------ %%
%% --------------------------------- PROOF PROPOSITION 3 ------------------------------------------------ %%
%% ------------------------------------------------------------------------------------------------------ %%
%% ------------------------------------------------------------------------------------------------------ %%
%% ------------------------------------------------------------------------------------------------------ %%

\bigskip 
\noindent

\textit{Proof of Proposition 3.} 

\textbf{Part 1.} \emph{Showing that the derivative of the vector of optimal performance-related compensation $\mathbf{v}^{*}$ with respect to link $g_{ij}$ is non-negative.} \medskip 

Recall 
\begin{equation*}
	\mathbf{v}^{*} = \frac{1}{1+c\eta\sigma^{2}}\mathbf{W}(\mathbf{G},\delta,\lambda) \boldsymbol{\alpha},
\end{equation*}
where $\alpha_{s} = \sum_{t=1}^{n}m_{ts}$, i.e., the total influence of agent $s$ on all other agents, including $s$. For the rest of the proof, I will omit the dependency on the networks $\mathbf{G}$ and parameters $\delta$ and $\lambda$ and simply write $\mathbf{W}$. As before, entry $(i,j)$th of $\mathbf{W}$ is denoted by $w_{ij}$.  

The derivative with respect to $g_{ij}$, using the chain rule, can be written as the sum of two terms:
\begin{equation}
	\frac{\partial\mathbf{v}^{*} }{\partial g_{ij}} = \frac{1}{1+c\eta\sigma^{2}}\left[\frac{\partial\mathbf{W} }{\partial g_{ij}}\cdot \boldsymbol{\alpha} + \mathbf{W} \cdot\frac{\partial \boldsymbol{\alpha}}{\partial g_{ij}}\right].
	\label{D_v_gij}
\end{equation}

Focusing on the first term in brackets, the derivative of $\mathbf{W}$ with respect to $g_{ij}$, which involves the derivative of the inverse of a matrix, is
\begin{equation*}
	\frac{\partial\mathbf{W} }{\partial g_{ij}} = -\mathbf{W} \frac{\partial [\mathbf{I}-\delta\mathbf{G}^{T}\mathbf{M}^{T}\mathbf{M}\mathbf{G}]}{\partial g_{ij}}\mathbf{W} = \delta \mathbf{W}\frac{\partial \mathbf{G}^{T}\mathbf{M}^{T}\mathbf{M}\mathbf{G}}{\partial g_{ij}}\mathbf{W}.
\end{equation*}

Applying the chain rule several times, the derivative above is given by
\begin{equation*}
	\frac{\partial \mathbf{G}^{T}\mathbf{M}^{T}\mathbf{M}\mathbf{G}}{\partial g_{ij}} = \left[ \frac{\partial \mathbf{G}^{T}}{\partial g_{ij}}\mathbf{M}^{T} + \mathbf{G}^{T}\frac{\partial \mathbf{M}^{T}}{\partial g_{ij}} \right] \mathbf{M}\mathbf{G} + \mathbf{G}^{T}\mathbf{M}^{T}\left[ \frac{\partial \mathbf{M}}{\partial g_{ij}} \mathbf{G} + \mathbf{M}\frac{\partial \mathbf{G}}{\partial g_{ij}} \right]. 
\end{equation*}

Then, the derivatives of $\mathbf{G}$ and $\mathbf{M}$ with respect to $g_{ij}$ are 
\begin{equation*}
	\frac{\partial \mathbf{G}}{\partial g_{ij}}= \mathbf{E}_{ij}, \quad \frac{\partial \mathbf{M}}{\partial g_{ij}}=\frac{\beta}{c}\mathbf{M}\mathbf{E}_{ij}\mathbf{M},
\end{equation*}
where $\mathbf{E}_{ij}$ is a matrix of zeros, except the $(i,j)$th entry which is equal to 1. Putting everything together, the first term in brackets of \cref{D_v_gij} is equal to
\begin{equation*}
	\frac{\partial\mathbf{W} }{\partial g_{ij}}\cdot \boldsymbol{\alpha} = \delta \mathbf{W} \left[ \mathbf{E}_{ji}\mathbf{M}^{T}\mathbf{M}\mathbf{G} + \frac{\beta}{c}\mathbf{G}^{T}\mathbf{M}^{T}\mathbf{E}_{ji}\mathbf{M}^{T}\mathbf{M}\mathbf{G} + \frac{\beta}{c}\mathbf{G}^{T}\mathbf{M}^{T}\mathbf{M}\mathbf{E}_{ij}\mathbf{M}\mathbf{G}+\mathbf{G}^{T}\mathbf{M}^{T}\mathbf{M}\mathbf{E}_{ij} \right] \mathbf{W}\cdot \boldsymbol{\alpha}\geq 0,
\end{equation*}
since all matrices are non-negative. 

Next, consider the second term in the brackets of \cref{D_v_gij}. The derivative of each entry of $\boldsymbol{\alpha}$ with respect to $g_{ij}$ is given by
\begin{equation}
	\frac{\partial \alpha_{s}}{g_{ij}} = \sum_{t=1}^{n}\frac{\partial m_{ts}}{\partial g_{ij}} = \sum_{t=1}^{n} \frac{\beta}{c} m_{ti}m_{js} = \frac{\beta}{c}m_{js}\sum_{t=1}^{n}m_{ti},
	\label{D_alphas_gij}
\end{equation}
where the second equality follows from taking the $(t,s)$ entry of the derivative of $\mathbf{M}$ with respect to $g_{ij}$ computed above. Note that this derivative is positive if $m_{js}>0$, i.e., if $j$ is influenced by $s$. Moreover, when $s=j$ this derivative is always strictly greater than zero. Hence, the second term in the brackets of \cref{D_v_gij} is 
\begin{equation*}
	\mathbf{W} \cdot\frac{\partial \boldsymbol{\alpha}}{\partial g_{ij}} = (\frac{\beta}{c}\sum_{t=1}^{n}m_{ti}) \mathbf{W} \cdot \begin{bmatrix}
		m_{j1} \\
		\vdots \\
		m_{jn}
	\end{bmatrix}\geq 0.
\end{equation*}

Thus, it is established that the derivative of $\mathbf{v}^{*}$ with respect to $g_{ij}$ is non-negative. \medskip 
 
\textbf{Part 2.} \emph{Show that for agents $k$ not weakly connected to $j$, the derivative of $v_{k}^{*}$ with respect to $g_{ij}$ is zero.} \medskip  

From part 1, the derivative of $v_{k}^{*}$ with respect to $g_{ij}$ can be written as
\begin{equation}
	\frac{\partial v^{*}_{k}}{\partial g_{ij}} = \frac{1}{1+c\eta\sigma^{2}} \sum_{l=1}^{n} \left[ \frac{\partial w_{kl}}{\partial g_{ij}} \cdot \alpha_{t} + w_{kl}\cdot \frac{\partial \alpha_{l}}{\partial g_{ij}} \right].
	\label{D_vk_gij}
\end{equation}

I need to show that entry $k$ of the derivative with respect to $g_{ij}$ is equal to zero when agent $k$ is not weakly connected to agent $j$. Consider an agent $k$ that is not weakly connected to agent $j$, i.e., when considering the network ignoring the directions of links, there is no path between $k$ and $j$. Without loss of generality, it is convenient to relabel agents so that those weakly connected to $j$ are $s\in\{1,2,\dots, i\}$, and those weakly connected to $k$ are  $r\in\{l,\dots,n\}$\footnote{Note that what matters for the partition is whether an agent is weakly connected to $j$ or not. The analysis for an agent $l$ that is not weakly connected to either $j$ or $k$ is identical to that of agent $k$.}. Note that $\mathcal{N} = \{1,2,\dots, i\} \cup \{j,k\} \cup \{l,\dots,n\} $. This relabeling allows us to treat $\mathbf{G}$ and $\mathbf{M}$ as block matrices given by
\begin{equation*}
	\mathbf{G} = \begin{pmatrix}
  {\mathbf{G_{j}}}  & \rvline & \mathbf{0}_j \\
\hline
  {\mathbf{0}}_{k} & \rvline & \mathbf{G_{k}}
\end{pmatrix} \quad \text{and} \quad \mathbf{M} = \begin{pmatrix}
  \mathbf{M_{j}}  & \rvline & \mathbf{0}_j \\
\hline
  \mathbf{0}_{k} & \rvline & \mathbf{M_{k}}
\end{pmatrix},
\end{equation*}
where $\mathbf{G}_{j}$ and $\mathbf{G}_{k}$ are $j\times j$ and $(n-j)\times (n-j)$ matrices that only involve direct and indirect neighbors of $j$ and $k$, respectively, $\mathbf{M}_{j}$ and $\mathbf{M}_{k}$ are $j\times j$ and $(n-j)\times (n-j)$ matrices that capture the account for the total weight of all walks in each weakly connected component, and $\mathbf{0}_{j}$ and $\mathbf{0}_{k}$ are $j\times (n-j)$ and $(n-j)\times j$ matrices of zeros. Notice that each non-zero block in $\mathbf{G}$ and $\mathbf{M}$ can have some zero entries and is not necessarily symmetric. 

Next, recall that $\mathbf{W}=\sum_{x=0}^{\infty}\delta^{x}((\mathbf{M}\mathbf{G})^{T}\mathbf{M}\mathbf{G})^{x}$. Thus, for $x=0$ we have that $\mathbf{W}=\mathbf{I}$, i.e., the diagonal entries $w_{ii}\geq 1$ by definition\footnote{This was also true for the diagonal entries of $\mathbf{M}(\mathbf{G},\frac{\beta}{c})$}. For any $x\geq 1$, the product $((\mathbf{M}\mathbf{G})^{T}\mathbf{M}\mathbf{G})^{x}$ results in a matrix with the same block-structure, i.e., blocks $(1,2)$ and $(2,1)$ are both zeros. Furthermore, the non-zero blocks of $((\mathbf{M}\mathbf{G})^{T}\mathbf{M}\mathbf{G})^{x}$ are symmetric\footnote{For a matrix $\mathbf{A}$, the products $\mathbf{A}\mathbf{A}^{T}$ and $\mathbf{A}^{T}\mathbf{A}$ gives a square and symmetric matrix, even if $\mathbf{A}$ is not square. In our case, $\mathbf{A}=(\mathbf{M}\mathbf{G})$.}. Thus, $\mathbf{W}$ can also be written as a block matrix:
\begin{equation*}
	\mathbf{W} = \begin{pmatrix}
  \mathbf{W_{j}}  & \rvline & \mathbf{0}_j \\
\hline
  \mathbf{0}_{k} & \rvline & \mathbf{W_{k}}
\end{pmatrix}
\end{equation*}
This implies that $w_{ks}=0$ for all $s\in\{ 1,\dots ,i\}\cup\{j\}$. Moreover, the derivative of $\alpha_{r}$ with respect to $g_{ij}$ (\cref{D_alphas_gij}) is always zero since $m_{jr}=0$ for all $r\in\{l,\dots, n \} \cup \{k\}$. That is, for all $s\in\{ 1,\dots ,i\}\cup\{j\}$, the second term in \cref{D_vk_gij} is zero by $w_{ks}=0$, and for all $r\in\{l,\dots, n \} \cup \{k\}$, the second term in \cref{D_vk_gij} is zero by $\partial \alpha_{r}/\partial g_{ij}=0$. Thus, the second term in \cref{D_vk_gij} is equal to zero for all $l\in \mathcal{N}$. 

Looking now at the first term in the brackets of \cref{D_vk_gij}, the fact that $w_{ks}=0$  for all $s\in\{ 1,\dots ,i\}\cup\{j\}$ implies that the derivative of these terms with respect to $g_{ij}$ will be zero for all $l=s$. I must show that the first term in brackets of \cref{D_vk_gij} is zero for all $l=r\in\{l,\dots,n\}\cup\{k\}$. To see this, recall that 
\begin{equation*}
	\frac{\partial\mathbf{W} }{\partial g_{ij}}\cdot \boldsymbol{\alpha} = \delta \mathbf{W} \left[ \mathbf{E}_{ji}\mathbf{M}^{T}\mathbf{M}\mathbf{G} + \frac{\beta}{c}\mathbf{G}^{T}\mathbf{M}^{T}\mathbf{E}_{ji}\mathbf{M}^{T}\mathbf{M}\mathbf{G} + \frac{\beta}{c}\mathbf{G}^{T}\mathbf{M}^{T}\mathbf{M}\mathbf{E}_{ij}\mathbf{M}\mathbf{G}+\mathbf{G}^{T}\mathbf{M}^{T}\mathbf{M}\mathbf{E}_{ij} \right] \mathbf{W}\cdot \boldsymbol{\alpha}.
\end{equation*}

We can write $\mathbf{E_{ij}}$ as a block matrix as follows:
\begin{equation*}
	\mathbf{E_{ij}} = \begin{pmatrix}
  {\mathbf{E}_{ij}}  & \rvline & {\mathbf{0}}_j \\
\hline
  {\mathbf{0}}_{k} & \rvline & {\mathbf{0}}_{kk}
\end{pmatrix},
\end{equation*}
where ${\mathbf{0}}_{kk}$ is a $(n-j)\times (n-j)$ matrix of zeros. Then, multiplying this matrix, or its transpose $\mathbf{E}_{ji}$, by the block matrices $\mathbf{M}$ or $\mathbf{M}^{T}$ will result in a block matrix with ${\mathbf{0}}_{kk}$ in the $(2,2)$ block. This means that the entire bracket in the equation above can be written as some matrix $\mathbf{H}$ with zero matrices in blocks (1,2), (2,1), and (2,2). Then, when multiplying $\mathbf{W}\cdot \mathbf{H}\cdot \mathbf{W}\cdot \boldsymbol{\alpha}$ we have 
\begin{equation*}
	\frac{\partial\mathbf{W} }{\partial g_{ij}}\cdot \boldsymbol{\alpha} = \delta \mathbf{W} \mathbf{H} \mathbf{W}\cdot \boldsymbol{\alpha} =\delta \begin{pmatrix}
  \mathbf{W_{j}} & \rvline & \mathbf{0}_j \\
\hline
  \mathbf{0}_{k} & \rvline & \mathbf{W_{k}}
  \end{pmatrix} \cdot \begin{pmatrix}
  \mathbf{H}  & \rvline & \mathbf{0}_j \\
\hline
  \mathbf{0}_{k} & \rvline & \mathbf{0}_{kk}
\end{pmatrix}\cdot \begin{pmatrix}
  \mathbf{W_{j}}  & \rvline & \mathbf{0}_j \\
\hline
  \mathbf{0}_{k} & \rvline & \mathbf{W_{k}}
  \end{pmatrix} \cdot \boldsymbol{\alpha} = \delta \begin{pmatrix}
  \mathbf{W}_{j}{\mathbf{H}}\mathbf{W}_{j}  & \rvline & {\mathbf{0}}_j \\
\hline
  \mathbf{0}_{k} & \rvline & \mathbf{0}_{kk}
\end{pmatrix} \cdot \boldsymbol{\alpha},
\end{equation*}
which implies that $\partial w_{kr}/\partial g_{ij}$ is equal to zero for all $r=\{l,\dots,n\}\cup\{k\}$. This establishes that an increase in $g_{ij}$ has no effect on the optimal performance-related compensation of agents that are not weakly connected to $j$. \medskip 

\textbf{Part 3.} \emph{Showing that for agents weakly connected to $j$, they must have at least one in-link for the derivative to be strictly positive.} \medskip 

Consider agent $s\in\{ 1,\dots ,i\}$ that is weakly connected to $j$,
\begin{equation*}
	\frac{\partial v^{*}_{s}}{\partial g_{ij}} = \frac{1}{1+c\eta\sigma^{2}} \sum_{l=1}^{n} \left[ \frac{\partial w_{sl}}{\partial g_{ij}} \cdot \alpha_{t} + w_{sl}\cdot \frac{\partial \alpha_{l}}{\partial g_{ij}} \right].
\end{equation*}

Notice that for $l=\{k,l,\dots, n\}$, $w_{sl}=0$. So, the only potential positive terms that remain are
\begin{equation}
	\frac{\partial v^{*}_{s}}{\partial g_{ij}} = \frac{1}{1+c\eta\sigma^{2}} \sum_{l\in\{1,\dots,s,\dots,j\}} \left[ \frac{\partial w_{sl}}{\partial g_{ij}} \cdot \alpha_{l} + w_{sl}\cdot \frac{\partial \alpha_{l}}{\partial g_{ij}} \right].
	\label{D_vs_gij}
\end{equation}

Suppose there is no $l\neq s$ such that $s$ has influence on $l$, i.e., $m_{ls}=0$ for all $l\neq s$. This implies that column $s$ of $\mathbf{G}_{j}$, and row $s$ of $\mathbf{G}^{T}$, is full of zeros. Moreover, column $s$ of $\mathbf{M}_{j}$, and row $s$ of $\mathbf{M}_{j}^{T}$, is full of zeros except for entry $(s,s)$ which is equal to one. For any $x\geq 1$, the product $((\mathbf{M}_{j}\mathbf{G}_{j})^{T}\mathbf{M}_{j}\mathbf{G}_{j})^{x}$ results in a matrix in which row $s$ is full of zeros. Therefore, row $s$ and column $s$ of $\mathbf{W}_{j}$ are full of zeros except for the shared entry $(s,s)$ which is equal to one, since for $x=0$ we have that $\mathbf{W}_{j}=\mathbf{I}_{j}$. That is, if $m_{ls}=0$ for all $l\in\{1,\dots,s,\dots,j\}$, then $w_{sl}=0$ for all $l\neq s$ and $w_{ss}=1$. Thus, \eqref{D_vs_gij} is equal to zero if agent $s$ has no influence on any other agent, i.e., there is no agent $s'$ fo which $g_{s's}>0$. 

Suppose that there is one $l\neq s$ such that $s$ has influence on $l$, i.e., $m_{ls}>0$. This implies that $l$ has no influence on any agent, i.e., $m_{s'l}=0$ for all $s' \in \{1, \dots, i,j\}$\footnote{If $l$ has an influence on other agents, then $s$ would also have an influence on them, and so there would only be additional walks to $s$.}. Then, column $s$ and $l$ of $\mathbf{G}_{j}$ are all zeros except for entry $(l,s)$. Because $l$ has no influence on any other agent, columns $s$ and $l$ of $\mathbf{M}_{j}$ are full of zeros except for entries $(s,s)$, $(l,s)$, and $(l,l)$. Notice that since $l$ has no influence on any other agent, we know that $w_{s'l}=w_{ls'}=0$ for all $s' \in \{1, \dots, i,j\}$ and $w_{ll}=1$. Because $s$ is weakly connected to $j$ and has an influence on $l$, it must be that there is some other agent $s''$ that has an influence on $l$. Thus, entry $w_{s,s''}$ will be positive. Since $s''$ is weakly connected to $j$, it must be that either $m_{js''}>0$ or that $m_{s''j}>0$ (or both). If the former is true, then $w_{ss''}>0$, and the second term in the brackets of \eqref{D_vs_gij} is positive for $l=s''$. If the latter is true, then $w_{sj}>0$, and the first term in the brackets of \cref{D_vs_gij} is positive for $l=j$. Either way, \cref{D_vs_gij} is strictly positive.

%% ------------------------------------------------------------------------------------------------------ %%
%% ------------------------------------------------------------------------------------------------------ %%
%% ------------------------------------------------------------------------------------------------------ %%
%% --------------------------------- PROOF PROPOSITION 4 ------------------------------------------------ %%
%% ------------------------------------------------------------------------------------------------------ %%
%% ------------------------------------------------------------------------------------------------------ %%
%% ------------------------------------------------------------------------------------------------------ %%

\bigskip 
\noindent

\textit{Proof of Proposition 4.} Take the derivative of the optimal performance-related compensation with respect to the strength of network effects: 

\begin{equation}
	\frac{\partial \mathbf{v}^{*}}{\partial \beta} = \frac{1}{c\eta\sigma^{2}} \left[ \frac{\partial \mathbf{\tilde{W}}}{\partial \beta} \boldsymbol{\alpha} +\mathbf{W}\frac{\partial \boldsymbol{\alpha}}{\partial \beta }\right]
	\label{A_P4_Derv}
\end{equation}

As before, the first term involves the derivative of the inverse of a matrix, so we use the identity of the derivative of inverted matrices to obtain
\begin{equation*}
	\frac{\partial \mathbf{W}}{\partial \beta} = \frac{1}{c\eta\sigma^{2}}\mathbf{W} \frac{\partial [\mathbf{M}^{T}\mathbf{M} - \mathbf{M} - \mathbf{M}^{T}]}{\partial \beta}\mathbf{W} =\frac{1}{c\eta\sigma^{2}} \mathbf{W}\left[  \frac{\partial \mathbf{M}^{T}}{\partial \beta}\mathbf{M} + \mathbf{M}^{T}\frac{\partial \mathbf{M}}{\partial \beta} - \frac{\partial \mathbf{M}^{T}}{\partial \beta} - \frac{\partial \mathbf{M}}{\partial \beta} \right]\mathbf{W}.
\end{equation*}

Next, note that
\begin{equation*}
	\frac{\partial \mathbf{M}}{\partial \beta} = \frac{1}{c}\mathbf{M}\mathbf{G}\mathbf{M},
\end{equation*}
so that 

\begin{equation}
	\begin{aligned}
		\frac{\partial \mathbf{W}}{\partial \beta}  & =\frac{1}{c^{2}\eta\sigma^{2}} \mathbf{W}\left[  \mathbf{M}^{T}\mathbf{G}^{T}\mathbf{M}^{T}\mathbf{M} + \mathbf{M}^{T}\mathbf{M}\mathbf{G}\mathbf{M} - \mathbf{M}^{T}\mathbf{G}^{T}\mathbf{M}^{T} - \mathbf{M}\mathbf{G}\mathbf{M} \right]\mathbf{W} \\
	& = \frac{1}{c^{2}\eta\sigma^{2}} \mathbf{W}[ \mathbf{M}^{T}\mathbf{G}^{T}\mathbf{M}^{T}(\mathbf{M}-\mathbf{I}) +(\mathbf{M}^{T}-\mathbf{I})\mathbf{M}\mathbf{G}\mathbf{M}]\mathbf{W} \geq 0
	\end{aligned}
	\label{A_P4_DerW}
\end{equation}
since all matrices are non-negative and the diagonal entries of $\mathbf{M}$ are at least 1. With strict inequality for agents with at least one link.

Next, take the derivative of $\alpha_{i}$ with respect to $\beta$:  
\begin{equation}
	\frac{\partial \alpha_{i}}{\partial \beta} = \frac{\partial \sum_{j=1}^{n} m_{ji}}{\partial \beta} = \sum_{j=1}^{n}\frac{\partial m_{ji}}{\partial \beta} = \frac{1}{c} \sum_{j=1}^{n} \left( \mathbf{M}\mathbf{G}\mathbf{M} \right)_{ji} \geq 0, 
	\label{A_P4_Deralpha}
\end{equation}
with strict inequality for agents with at least one in-link. Note that \eqref{A_P4_DerW} and \eqref{A_P4_Deralpha} imply that \eqref{A_P4_Derv} is non-negative. Moreover, the derivative is strictly positive for agents with at least one in-link. The result for equilibrium efforts follows from the derivatives above.

%% ------------------------------------------------------------------------------------------------------ %%
%% ------------------------------------------------------------------------------------------------------ %%
%% ------------------------------------------------------------------------------------------------------ %%
%% --------------------------------- PROOF PROPOSITION 5 ------------------------------------------------ %%
%% ------------------------------------------------------------------------------------------------------ %%
%% ------------------------------------------------------------------------------------------------------ %%
%% ------------------------------------------------------------------------------------------------------ %%

\bigskip 
\noindent

\textit{Proof of Proposition 5.} We compute the derivative of the optimal performance-related compensation with respect to the cost of effort:

\begin{equation}
	\frac{\partial \mathbf{v}}{\partial c} = \frac{-\eta\sigma^{2}}{(1+c\eta\sigma^{2})^{2}}\mathbf{W}\alpha + \frac{1}{1+c\eta\sigma^{2}}\left[ \frac{\partial \mathbf{W}}{\partial c}\boldsymbol{\alpha} + \mathbf{W}\frac{\partial \boldsymbol{\alpha}}{\partial c} \right]
	\label{A_P5_Derv}
\end{equation}

First, 
\begin{equation}
	\frac{\partial \mathbf{W}}{\partial c} = \mathbf{W}\frac{\partial \left[\delta (\mathbf{M}\mathbf{G})^{T}\mathbf{M}\mathbf{G} \right]}{\partial c} \mathbf{W} = \mathbf{W}\frac{\partial \left[\frac{\partial \delta }{\partial c}(\mathbf{M}\mathbf{G})^{T}\mathbf{M}\mathbf{G} + \delta \frac{\partial \left( \mathbf{G}^{T}\mathbf{M}^{T}\mathbf{M}\mathbf{G} \right)}{\partial c} \right]}{\partial c} \mathbf{W}
	\label{A_P5_DerW}
\end{equation}
where 
\begin{equation*}
	\frac{\partial \delta}{\partial c} = \frac{-c(2+3c\eta\sigma^{2})}{(c^{2}(1+c\eta\sigma^{2})^{2}}\leq 0;
\end{equation*}
and 
\begin{equation*}
	\frac{\mathbf{G}^{T}\mathbf{M}^{T}\mathbf{M}\mathbf{G} }{\partial c} = \mathbf{G}^{T}\left(\frac{\partial \mathbf{M}^{T}}{\partial c}\mathbf{M} + \mathbf{M}^{T}\frac{\partial \mathbf{M}}{\partial c}\right)\mathbf{G}=\mathbf{G}^{T}\left(-\frac{\beta}{c^{2}}\mathbf{M}^{T}\mathbf{G}^{T}\mathbf{M}^{T}\mathbf{M}-\frac{\beta}{c^{2}}\mathbf{M}^{T} \mathbf{M}\mathbf{G}\mathbf{M}\right)\mathbf{G}\leq 0.
\end{equation*}

Second, 
\begin{equation}
	\frac{\partial \alpha_{i}}{\partial c} = \frac{\partial \sum_{j=1}^{n} m_{ji}}{\partial c} = \sum_{j=1}^{n}\frac{\partial m_{ji}}{\partial c} = -\frac{\beta}{c^{2}} \sum_{j=1}^{n} \left( \mathbf{M}\mathbf{G}\mathbf{M} \right)_{ji} \leq 0, 
	\label{A_P5_Deralpha}
\end{equation}

Thus, by \eqref{A_P5_DerW} and \eqref{A_P5_Deralpha}, we have that \eqref{A_P5_Derv} is non-positive. The result for the equilibrium of effort follows from the derivatives above. 

Next, we compute the derivative of the optimal performance-related compensation with respect to the coefficient of absolute risk aversion:

\begin{equation}
	\frac{\partial \mathbf{v}}{\partial \eta} = \frac{-c\sigma^{2}}{(1+c\eta\sigma^{2})^{2}}\mathbf{W}\alpha + \frac{1}{1+c\eta\sigma^{2}}\left[ \frac{\partial \mathbf{W}}{\partial \eta}\boldsymbol{\alpha} + \mathbf{W}\frac{\partial \boldsymbol{\alpha}}{\partial \eta} \right].
	\label{A_P6_Derv}
\end{equation}

We have that 
\begin{equation}
	\frac{\partial \mathbf{W}}{\partial \eta} = \mathbf{W}\frac{\partial \left[\delta (\mathbf{M}\mathbf{G})^{T}\mathbf{M}\mathbf{G} \right]}{\partial \eta} \mathbf{W} = \mathbf{W}\frac{\partial \left[\frac{\partial \delta }{\partial \eta}(\mathbf{M}\mathbf{G})^{T}\mathbf{M}\mathbf{G} + \delta \frac{\partial \left( \mathbf{G}^{T}\mathbf{M}^{T}\mathbf{M}\mathbf{G} \right)}{\partial \eta} \right]}{\partial \eta} \mathbf{W}
	\label{A_P6_DerW}.
\end{equation}
The derivative of $\mathbf{G}^{T}\mathbf{M}^{T}\mathbf{M}\mathbf{G}$ with respect to $\eta$ is zero. So we only need
\begin{equation*}
	\frac{\partial \delta}{\partial \eta} = -\frac{c^{3}\sigma^{2}}{(c^{2}(1+c\eta\sigma^{2})^{2}}\leq 0
\end{equation*} 
to conclude that \eqref{A_P6_Derv} is non-positive, since the last derivative in \eqref{A_P6_Derv} is also zero. The derivative with respect to the variance of the output shock is similar. 
%% ------------------------------------------------------------------------------------------------------ %%
%% ------------------------------------------------------------------------------------------------------ %%
%% ------------------------------------------------------------------------------------------------------ %%
%% ------------------------------------------------- END ------------------------------------------------ %%
%% ------------------------------------------------------------------------------------------------------ %%
%% ------------------------------------------------------------------------------------------------------ %%
%% ------------------------------------------------------------------------------------------------------ %%

\end{document}